\documentclass[aps,preprint,showpacs]{revtex4}

\usepackage{graphicx}
\usepackage{hyperref}
\usepackage{amsfonts}
\usepackage{amsbsy}
\usepackage{amssymb}
\usepackage{xcolor}
\usepackage[normalem]{ulem}
\begin{document}

\title{Statistical characterization of discrete conservative systems: \\
The web map}

\author{Guiomar Ruiz$^{1,3}$}
\email{guiomar.ruiz@upm.es}
\author{Ugur Tirnakli$^{2,3}$}
\email{ugur.tirnakli@ege.edu.tr}
\author{Ernesto P. Borges$^{4,5}$}
\email{ernesto@ufba.br}
  \author{Constantino Tsallis$^{3,5,6,7}$}
\email{tsallis@cbpf.br}

\affiliation{
$^1$Departamento de Matem\'{a}tica Aplicada y Estad\'{\i}stica, Universidad Polit\'{e}cnica de Madrid,
Pza.\ Cardenal Cisneros s/n, 28040 Madrid, Spain \\
$^2$Department of Physics, Faculty of Science, Ege University, 35100 Izmir, Turkey \\
$^3$Centro Brasileiro de Pesquisas Fisicas \\ \mbox{Rua Xavier Sigaud 150, Rio de Janeiro 22290-180, Brazil}\\
$^4$Instituto de F\'isica, Universidade Federal da Bahia, Salvador-BA 40170-115 Brazil \\
$^5$National Institute of Science and Technology for Complex Systems \\
\mbox{Rua Xavier Sigaud 150, Rio de Janeiro 22290-180, Brazil} \\
 $^6$ Santa Fe Institute, 1399 Hyde Park Road, Santa Fe, New Mexico 87501, USA \\
 $^7$ Complexity Science Hub Vienna, Josefst\"adter Strasse 39, 1080 Vienna, Austria
 }

\date{\today}

\begin{abstract}
We numerically study the two-dimensional, area preserving, web map. When the map is
governed by ergodic behavior, it is, as expected, correctly described by Boltzmann-Gibbs statistics,
based on the additive entropic functional $S_{BG}[p(x)] = -k\int dx\,p(x) \ln p(x)$. In contrast,  possible ergodicity
breakdown and transitory sticky dynamical behavior drag the map into the realm of generalized $q$-statistics,
based on the nonadditive entropic functional $S_q[p(x)]=k\frac{1-\int dx\,[p(x)]^q}{q-1}$
($q \in {\cal R}; S_1=S_{BG}$). We statistically describe the system (probability distribution of the sum of successive iterates, sensitivity to the initial condition, and
entropy production per unit time)
for typical values of the parameter that controls the ergodicity of the map. For small (large)
values of the external parameter $K$, we observe $q$-Gaussian distributions with $q=1.935\dots$
(Gaussian distributions), like for the standard map. In contrast, for intermediate values of $K$,
we observe a different scenario, due to the  fractal structure of the trajectories embedded in the chaotic sea. Long-standing non-Gaussian distributions are characterized in terms of the kurtosis and the box-counting dimension of  chaotic sea.
\end{abstract}

\maketitle

\section{Introduction}
\label{sec:1}
As well--known, invariant closed curves of area--preserving maps present complete barriers to orbits
evolving inside resonance  islands in the two--dimensional phase space. Outside these regions, there exist
families of smaller islands and invariant Cantor sets, to which chaotic orbits are observed to ``stick'' for very
long times. Thus, at the boundaries of these islands, an ``edge of chaos'' develops with vanishing or very small
Lyapunov exponents, where trajectories yield quasi-stationary states (QSS) that are often very long--lived.
Such phenomena have been thoroughly studied to date in terms of a number of \textit{dynamical}
mechanisms responsible for chaotic transport in area--preserving maps and low--dimensional Hamiltonian
systems \cite{MacKay, Wiggins}.

In such a weakly chaotic regime, chaotic orbits ergodically wander through a subset
of the energy surface without ever covering it completely, and  ``islands of stability'' are associated with stable
periodic orbits that are caused by invariant curves encircling stable periodic points that exclude the
surrounding chaotic trajectory. There are also many island chains that correspond to orbits of different
period and a hierarchy of stable points with island chains surrounding island chains {\it ad infinitum}.
This hierarchical organization causes the surrounding chaotic orbit to have structure at all scales.

A distinctive
feature of all fractals is the dependence of the apparent size on the scale of resolution. In this line of approach, Umberger and
Farmer characterize chaotic orbits in a two dimensional conservative system as
{\it fat fractals} that have positive Lebesgue measure but their apparent size depends on the
scale of resolution \cite{Umberger85,Farmer85}.  Of course, an orbit is composed of a countable set of points
and has no area, but when we refer to the ``area of an orbit'' we actually mean the Lebesgue measure of the
closure of the orbit.
Consequently, we can characterize chaotic orbits of a map as {\it fractals} if the apparent area occupied by the orbit
depends on the resolution used to measure it. However, it is convenient to distinguish the ``fat'' fractals from ``thin''
(not {\it fat}) fractals of zero Lebesgue measure, such as the strange attractors that appear in dissipative
dynamics \cite{Cantorlike}. In fact, the existence of
disjoint invariant regions with a different degree of stochasticity on the same constant energy surface
has been investigated  by  Pettini and Vulpiani in nonlinear hamiltonian systems, and  their results do suggest fractal dimensions  of the subspaces spanned by the trajectories \cite{Pettini84}.

On the other hand, Benettin et al. have investigated the dimensionality of {\it one} finite-time trajectory
near the unstable manifold of a family of two-dimensional perturbed integrable area preserving maps.
They conclude that a finite-time trajectory will necessarily exhibit an apparent fractal dimension, which will be the effective one to all practical purposes: in order to find an effective dimension $d_f=2$, one has to look at
sufficiently small scales that decrease exponentially fast with the inverse of the parameter of the map  \cite{Benettin86}.

In this work, we shall consider a particular two-dimensional area-preserving map whose sticky behavior appears to play a significative role in the dynamics, namely, the web map  \cite{Zaslavskii}. The previous considerations make  the complexity of this map  susceptible to be studied in the context of
generalized $q$-statistics \cite{Tsallis,Tsallis2010}.
According to this approach, based on the nonadditive entropy $S_q$, whose formulation was inspired in the geometry of multifractals, the probability density functions (pdfs) that
optimize  $S_q$  --- under appropriate constraints --- are $q'$-Gaussian distributions that represent
metastable states or QSS of the dynamics.
Generalized $q$-statistics manages to characterize meta-stable or stationary states by a triplet of
$q$-values, i.e., the $q$-triplet $(q_{sens}, q_{rel}, q_{stat})$, where {\it sens } stands for {\it sensitivity},
{\it rel } stands for {\it relaxation} and {\it stat } stands for {\it stationary}, whose values collapse to unity when
ergodicity is attained (i.e., $q_{sens} = q_{rel} = q_{stat} = 1$). In fact, in the case of ergodicity, the
Boltzmann-Gibbs entropy is the proper one ($q_{sens} = 1$), and Gaussian distributions are observed
($q_{stat} = 1$).

In this scenario, we will analyze the  dependence of the apparent size on the scale of resolution --- the capacity  dimension ---  of the subspace spanned by a set of finite-time trajectories embedded inside the chaotic sea of the web map. This is to reveal, for some paradigmatic values of the parameter of the map, the relation between a fractal dimension of the subspaces spanned by the trajectories embedded inside the chaotic map and their respective pdfs. We will also analyze other parameters that characterize the nonextensivity of the map.

The paper is organized as follows.
In Section~\ref{sec:2}  we present
the area--preserving web map,   describing  the role of the parameter of the map in the process of ergodicity breakdown and the appearance of   QSS. In Section~\ref{sec:3} we analyze the probability distribution of the sum of the iterates of the map, and  we exhibit that  the fractal dimension of the trajectories embedded inside the  apparently chaotic sea ---  even in the limit of an infinite number of iterations --- appears to be a sufficient condition for the convergence to non-Gaussian distributions.  In Section~\ref{sec:4} we analyze  non-extensive indices related to the sensitivity to the initial conditions and the
entropy production per unit time. Our main conclusions are drawn in Section~\ref{sec:5}.

\section{Ergodicity breakdown }
\label{sec:2}
The web map is defined  as:
\begin{equation}
\begin{array}{l}
u_{i+1}=v_i \\v_{i+1}=-u_i-K \sin (v_i)
\end{array}
\label{eq:map}
\end{equation}
where both $u_n$ and $v_n$ are taken as modulo $2 \pi$, and $K$ is the parameter  that controls the ergodicity of the map. When $K=0$, the system is integrable and all orbits are $T=4$-periodic. Increasing  $K$,  invariant orbits that correspond to periodic motion disappear, and stable elliptic periodic points, unstable hyperbolic periodic points and chaotic orbits, appear in their wake.
Tuning up the value of  $K$,
the phase portraits exhibit a clear evolution from a predominance of stability islands all over the phase space
(e.g., $K=0.1$) to the flood of a whole chaotic sea that occupies the whole square (e.g.. $K=5.0$). The phase
portraits of intermediate cases (e.g., $K=3$, $K=3.5$, $K=3.7$, $K=3.8$, $K=4.5$...) show the stability islands and the chaotic sea
coexisting on the full phase space of the map.

\begin{figure}[ht]
\centering
\includegraphics[height=6cm,angle=0]{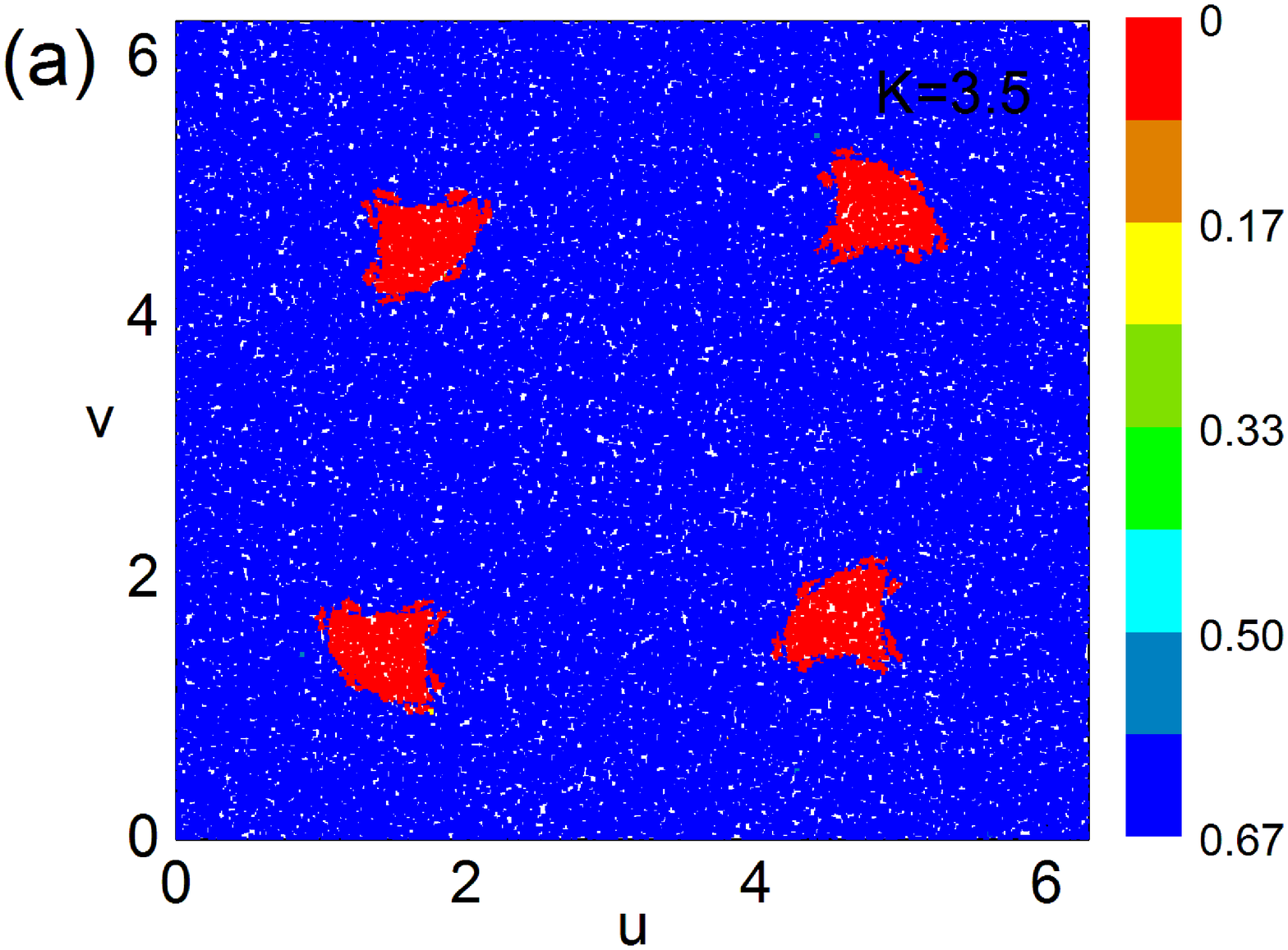}
\includegraphics[height=6cm,angle=0]{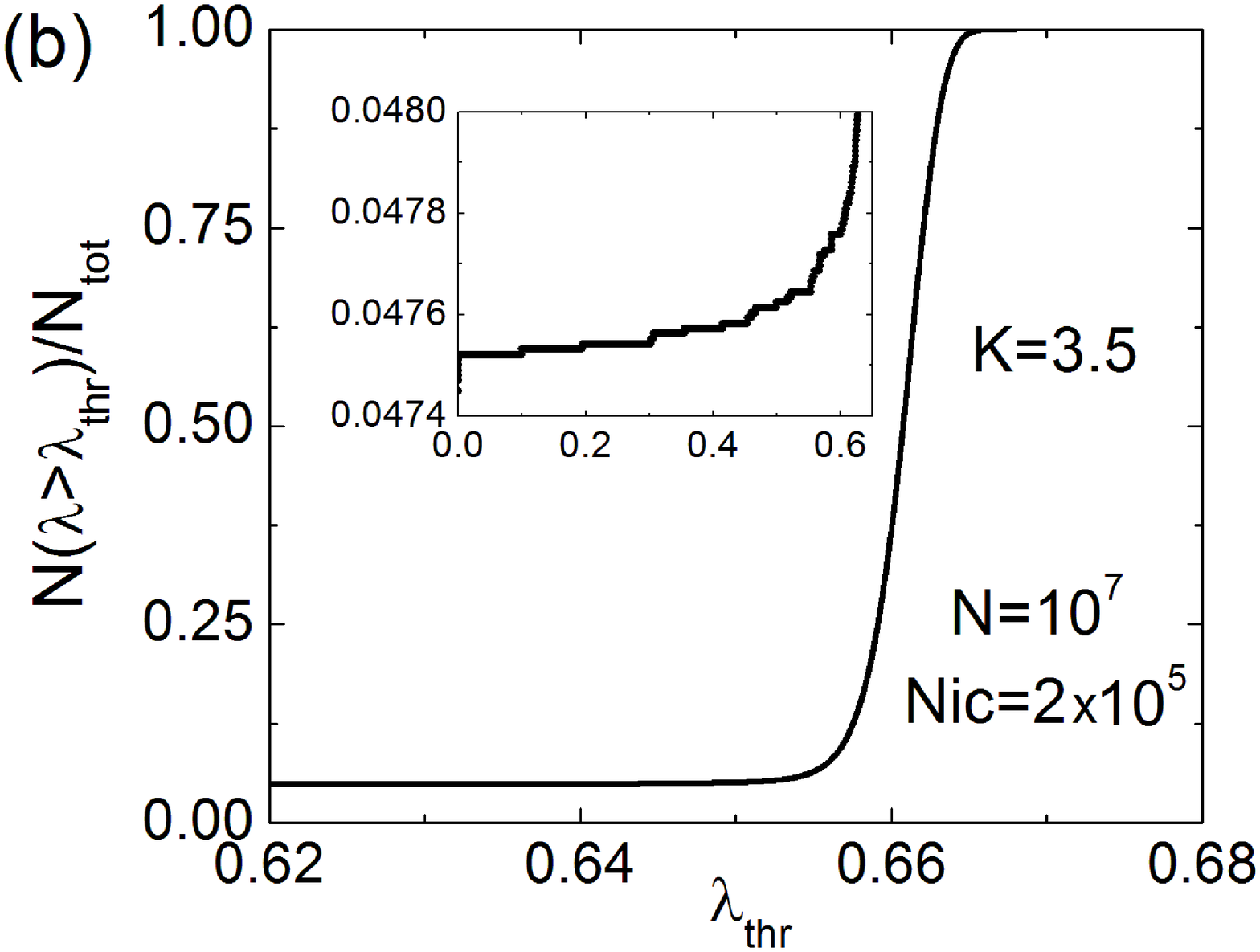}
\vspace{-0.4cm}
\caption{\label{fig:lyap}
{\small (Color online) (a) Finite-time largest Lyapunov exponent (LLE) of $2 \times 10^5$ randomly
chosen initial conditions of the $K=3.5$ web map,  for $10^7$ iterations. The Lyapunov spectrum
presents largely positive LLE where the phase space is dominated by a chaotic sea, and  nearly vanishing
LLE in four regions where stability islands dominate the dynamical behavior. (b)  Cumulative
distribution of the relative number of initial conditions, as a function of a  finite-time LLE
threshold, $\lambda_{thr}$. }}
\end{figure}

For fixed values of $K$, and using Benettin algorithm \cite{Benettin}, we calculate the Largest Lyapunov Exponent (LLE)  for each initial
condition separately --- or, alternatively, the Smaller Alignment Index (SALI)--- to characterize the orbits and the possible existence of QSS \cite{Skokos}.
In fact, the finite-time contributions coming from the initial conditions of stability islands differ considerably
from the ones coming from the strongly chaotic sea.
Fig.~\ref{fig:lyap} shows, for a representative value of the map parameter that induces the
coexistence of stability islands and chaotic sea, i.e., $K=3.5$, the large but finite-time LLEs that have been
obtained for a huge random set of initial conditions all over the phase space. The cumulative distribution
of the relative number of initial conditions, as a function of the finite-time LLE threshold, presents an abrupt
increase and  suggests that  we must take into account only two sets of trajectories in the statistics:
the strongly chaotic trajectories embedded in a chaotic sea whose LLE $\gg 0$, and the
weakly chaotic trajectories whose LLE $\simeq 0$. If we compare this scenario with that one found in the standard map  \cite{Ugur}, an analogous statistical behavior would be expected and, consequently, the coexistence of two different regimes
would induce a linear superposition of their respective Gaussian and $q$-Gaussian probability distribution functions. But it is not so, as we will show in Section~\ref{sec:3}.

\begin{figure}[h!]
\centering
\includegraphics[height=7cm,angle=0]{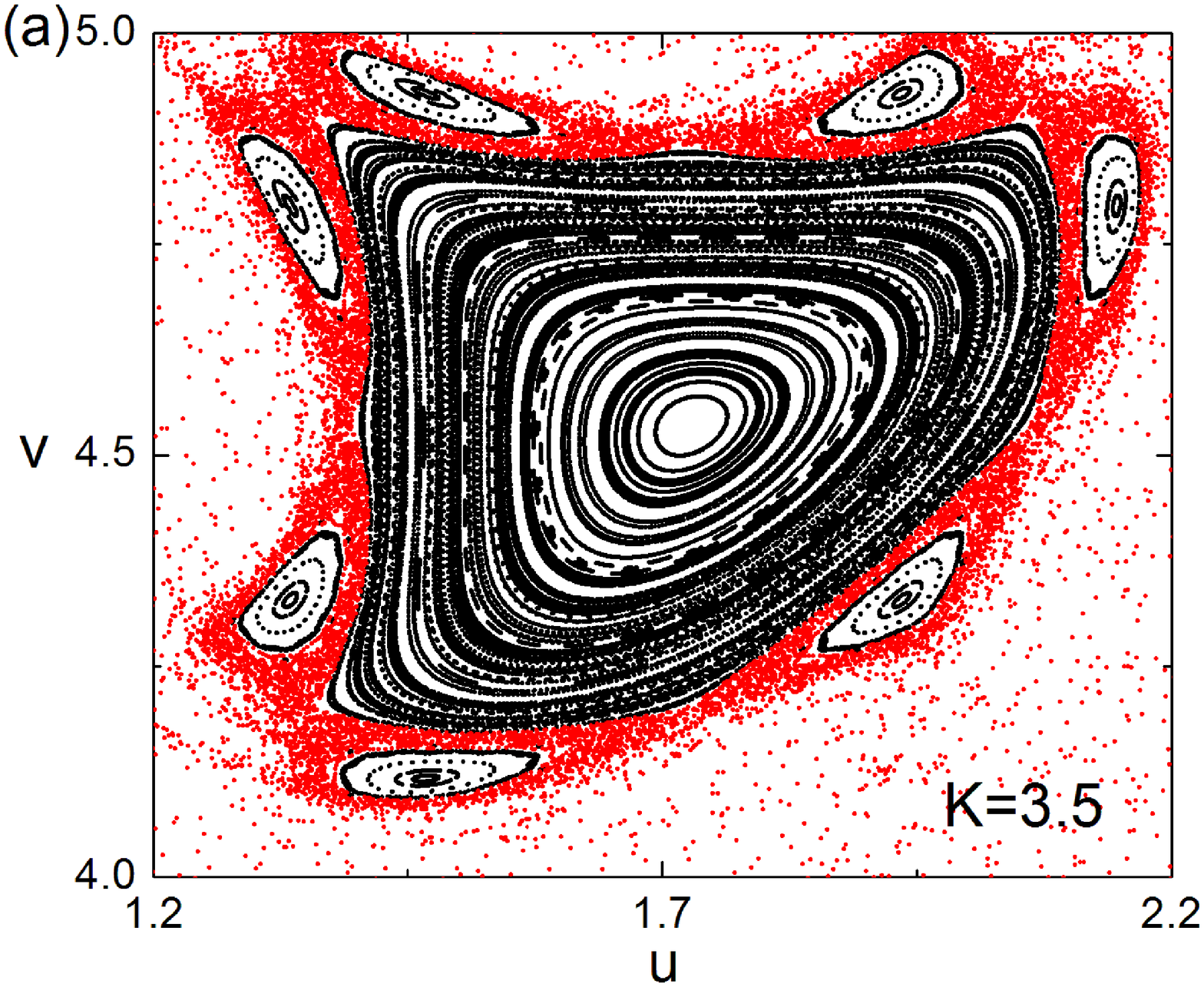}
\includegraphics[height=7cm,angle=0]{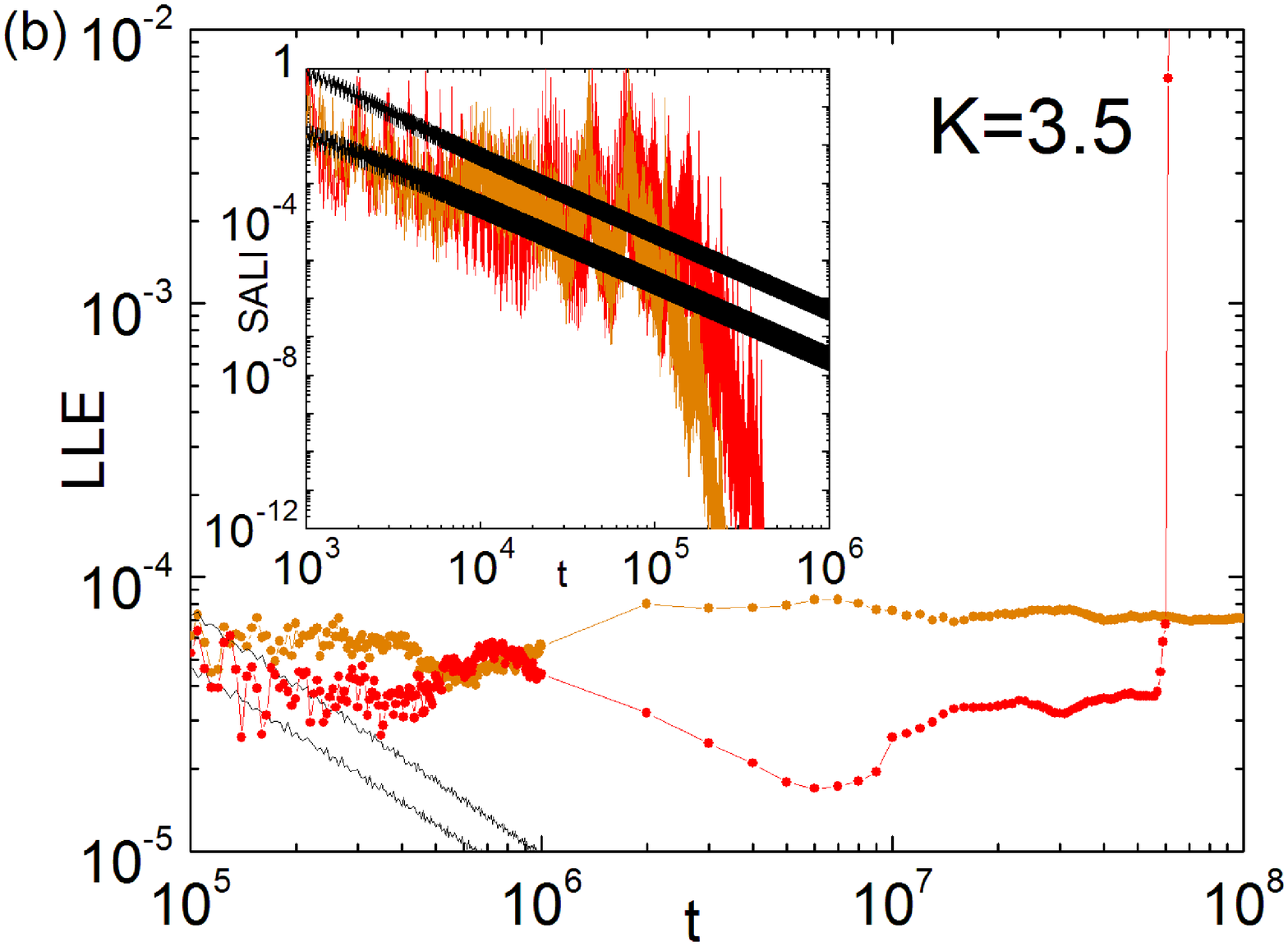}
\vspace{-0.4cm}
\caption{\label{fig:stikytray}
{\small (Color online) (a) A zoom of the phase space of the web map for $K=3.5$, that shows sticky behavior
(red trajectories) around quasiperiodic islands (black trajectories). (b) Finite-time LLE and SALI evolution of
two quasiperiodic orbits (black), and two strongly chaotic orbits with sticky behavior: one of them (red)
{\it escapes} after $6 \times 10^7 $ iterations, and the other one (orange) does not escape before $10^8$
iterations. }}
\end{figure}

On the other hand, it seems that for some values of the external parameter of the web map --- and, in particular, for $K=3.5$ --- the sticky effect of strongly chaotic trajectories
appears to be statistically much more significant than in the standard map. Fig.~\ref{fig:stikytray} exemplifies
 trajectories with extremely small finite-time LLE that suddenly escape from sticky regions to larger
finite-time LLE at  larger times. The inverse phenomenon also occurs, as trajectories initially embedded in a
strongly chaotic sea (LLE~$\gg 0$) can be trapped around the islands (LLE~$\sim 0$), after an arbitrarily large
time evolution.

Let us now show how ergodicity breakdown
and the described transitory sticky
dynamical behavior  drag the map into the realm of a generalized statistics.

\section{Stationary and quasistationary distributions}
\label{sec:3}
It is well known, through the central limit theorem, that in case of trajectories which are
essentially ergodic and mixing, Gaussians are ultimately observed as the probability density
distributions of the sums of the iterates of the map. In such cases, the LLE is bounded away
from zero. On the contrary, in the case of vanishing LLE, it has been observed that the
re-scaled sums are not Gaussians but  can instead appear to approach $q$-Gaussian limit distributions.
This is in fact the case of the  standard map \cite{Ugur,Ruiz} and, for some $K$ values, it is
also the case of the web map. But we have also found a third unexpected scenario, for a kind
of trajectories  embedded inside the chaotic sea ---  whose LLE is consequently  bounded
away from zero ---,  that specially distinguishes the  statistical behavior of the web map from
that of the standard map. Let us now describe in detail these three distinct  scenarios.

In the spirit of the Central Limit Theorem, let us  define the variable

\begin{equation}
y= \sum_{i=1}^{N} \left(u_i- \langle   \,  u  \, \rangle  \right)
\label{eq:variable}
\end{equation}
where $\langle   \,  \cdots \rangle$ implies averaging over a large number of iterations $N$ and
a large number of randomly chosen initial conditions $M$, i.e.,
$\langle  u \rangle=\frac{1}{M}\frac{1}{N}\sum_{j=1}^{M}\sum_{i=1}^N u_i^{(j)}$. It was
previously shown, for arbitrary values of the parameter $K$ of the standard
map \cite{Ugur,Ruiz}, that the probability distribution of these sums (Eq.~(\ref{eq:variable})) can be
modeled as

\begin{equation}
 \label{eq:qGauss}
P_q(y;\mu_q,\sigma_q)=A_q\sqrt{B_q}\left[1-(1-q)B_q(y-\mu_q)^2
\right]^{\frac{1}{1-q}},
\end{equation}
that represents the probability density for the initial conditions inside the vanishing Lyapunov
region ($q\ne1$), where $\mu_q$ is the $q$-mean value,  $\sigma_q$ is the $q$-variance,
$A_q$ is the normalization factor and $B_q$ is a parameter which characterizes the width of
the distribution \cite{prato-tsallis-1999}:

\begin{equation}
 \label{eq:Aq}
 A_q=\left\{\begin{array}{lc}\displaystyle\frac{\Gamma\left[\frac{5-3q}{2(1-q)}\right]}
 {\Gamma\left[\frac{2-q}{1-q}\right]}\sqrt{\frac{1-q}{\pi}}, &q<1\\
  \displaystyle\frac{1}{\sqrt{ \pi }},&q=1\\
\displaystyle\frac{\Gamma \left[\frac{1}{q-1}\right]}{\Gamma\left[\frac{3-q}{2(q-1)}\right]}
\sqrt{\frac{q-1}{\pi}}, &1<q<3
\end{array} \right.
\end{equation}

\begin{equation}
 \label{eq:Bq}
 B_q=[(3-q)\sigma_q^2]^{-1} .
\end{equation}
The $q$-mean value and $q$-variance are defined by
(see  \cite{prato-tsallis-1999} for the continuous version):
\begin{equation}
\mu_q = \frac{ \sum_{i=1}^N y_i [P_q(y_i)]^q }
              { \sum_{i=1}^N     [P_q(y_i)]^q },
\end{equation}

\begin{equation}
\sigma_q^2 = \frac{ \sum_{i=1}^N y_i^2 [P_q(y_i)]^q }
                 { \sum_{i=1}^N       [P_q(y_i)]^q },
\end{equation}
though we have considered these variables as fitting parameters.

\begin{figure}[h!]
\centering
\includegraphics[height=7cm,angle=0,clip=]{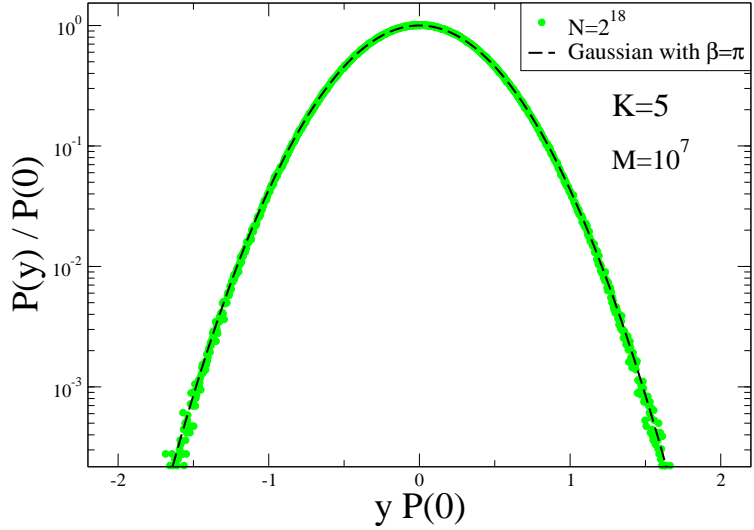}\vspace{-0.4cm}
\caption{\label{fig:CLTGaussian}
{\small (Color online) The probability distribution function of the sum of the iterates of $K=5$
web map (green dots) neatly fits a Gaussian (dashed line), $P(y)=P(0)e^{-\beta y^2}$
 where $\beta=3.14$. $N$ and $M$ are the number of iterates and initial conditions,
 respectively. }}
\end{figure}

The $q \to 1$ limit recovers the Gaussian distribution
$P_1(y; \mu_1, \sigma_1)=\frac{1}{\sigma_1 \sqrt{2 \pi }}
\exp\left[-\frac{1}{2}\left(\frac{y-\mu_1}{\sigma_1}
\right)^2\right]$ and, in fact,
Fig.~\ref{fig:CLTGaussian} shows that, when the trajectories of the web map are essentially ergodic and mixing
(LLE is bounded away from zero and a chaotic sea involves the whole phase portrait,
e.g., $K=5$), the limit probability distribution of
Eq.~(\ref{eq:variable})
 neatly approaches a Gaussian
even for a relatively small number of iterations $N$.

\begin{figure}[h!]
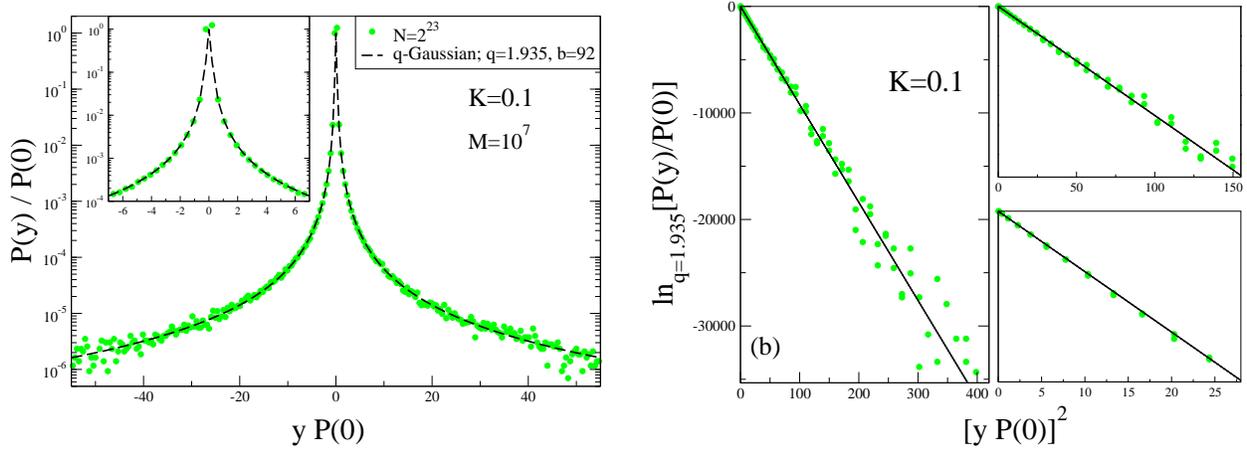

\centering
\includegraphics[width=0.48\columnwidth,keepaspectratio,clip=]{fig4a.eps}\hspace{0.5cm}
\includegraphics[width=0.48\columnwidth,keepaspectratio,clip=]{fig4b.eps}\vspace{-0.4cm}
\caption{\label{fig:CLTqGaussian}
{\small (Color online) (a)~$K=0.1$ probability distribution function $P(y)$ (green dots)
demonstrates to fit a $q$-Gaussian (dashed line), i.e., $P(y)=P(0)e_q^{-\beta y^2}$, where
$q=1.935$ and $\beta=92$. $N=2^{23}$ and $M=10^7$ are the number of iterates and initial
conditions, respectively. (b)~$q$-logarithmic representation of the same distribution is given
for the tail, intermediate and central regions.}}
\end{figure}

On the contrary, when the phase space portrait of the web map is dominated by the stability islands,
the probability  distribution of
Eq.~(\ref{eq:variable}) converges to a $q$-Gaussian with $q \simeq 1.935$, as shown in Fig.~\ref{fig:CLTqGaussian}.
This convergence has also been observed for values of the
map parameter sufficiently close to $K=0$.
The persistence of the value  $q_{stat}=1.935$ for both the
standard and the web maps, when  space portrait maps are dominated by stability islands,  constitutes
an intriguing result.

On the other hand, intermediate values of $K$, where both chaotic sea and stability islands coexist, appear to
confirm that the probability distribution all over the whole phase space of the map is well-fitted by a superposition of a
$(q=1.395)$-Gaussian and another distribution. But this case is much more intricate than
that of the standard map, and hereinafter we will restrict our analysis to the distributions of
trajectories embedded inside the chaotic sea, where we have obtained some astonishing results.

\begin{figure}[h!]
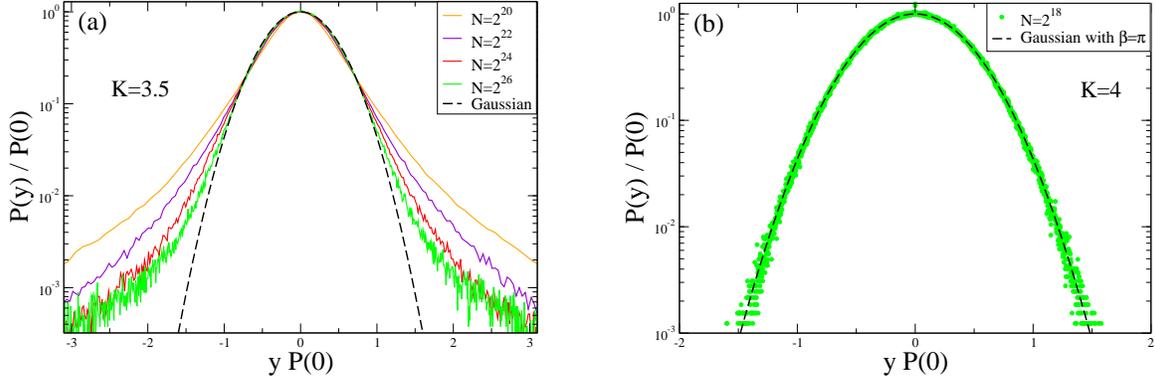

\centering
\includegraphics[height=5cm,angle=0,clip=]{fig5a.eps}\hspace{1cm}
\includegraphics[height=5cm,angle=0,clip=]{fig5b.eps}
\caption{\label{fig:HistgK35}
{\small
(Color online) (a) Probability distribution function for the iterates of $K=3.5$ web map for
various $N$ values. The number of iterations is $M=10^7$ all taken from the chaotic sea.
(b) The same for $K=4$.}}
\end{figure}

Actually, an unexpected result is that, for particular map parameter values, the
probability distribution of the finite trajectories embedded inside the chaotic sea appear to be
well far away from a Gaussian even for extremely long iteration times. For relatively large
number of iterations ($N\sim 2^{22}$), the probability distribution appears to be well fitted by
a $q$-Gaussian but, for even larger iterations, the central part appears to evolve in an
extremely slowly rhythm  towards a Gaussian, as shown in Fig.~\ref{fig:HistgK35}, but non-Gaussian
tails do persist.

\begin{figure}[h!]
\hspace{0.3cm}
\includegraphics[height=5cm,angle=0,clip=]{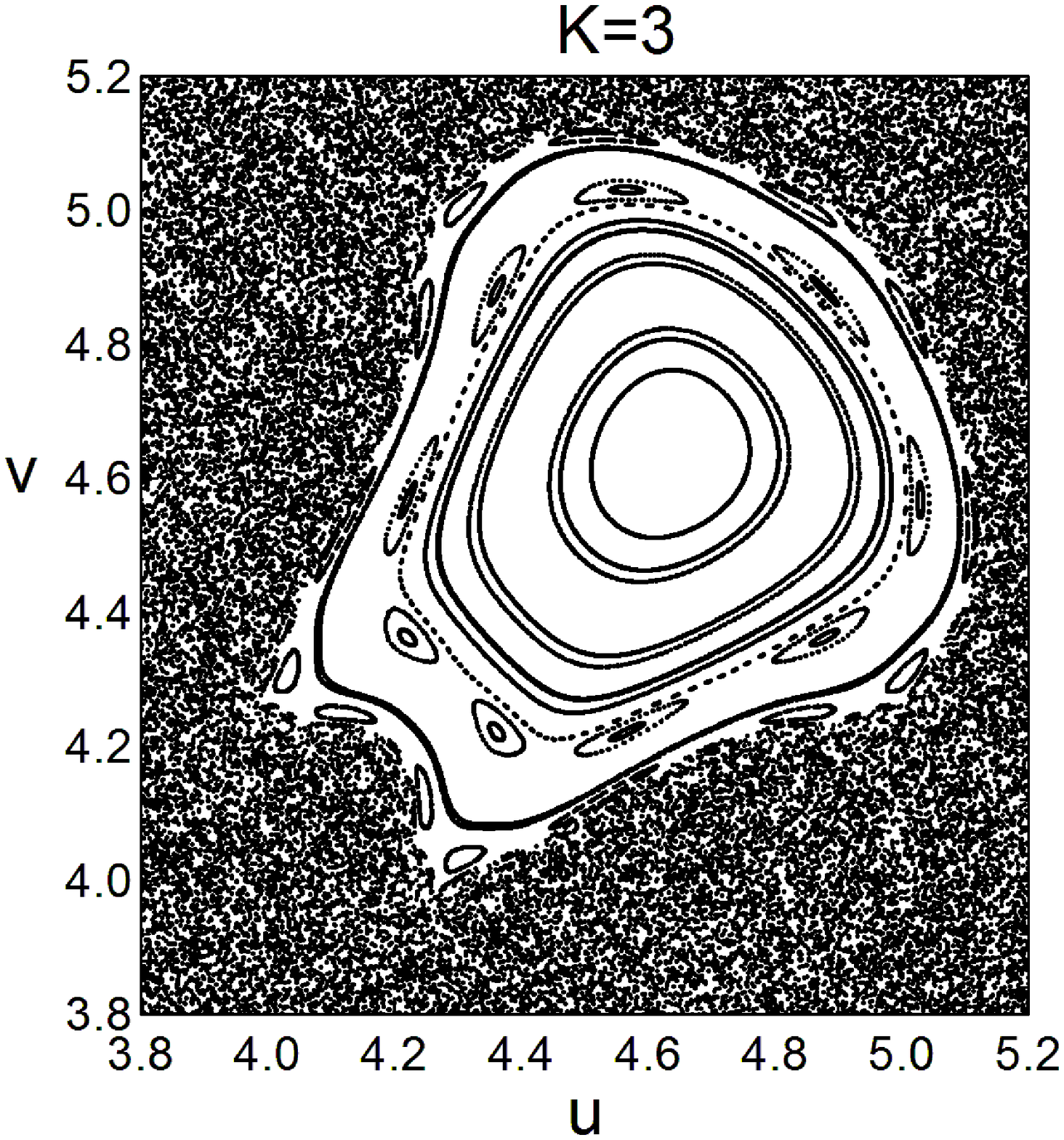}
\hspace{-0.3cm}
\includegraphics[height=5cm,angle=0,clip=]{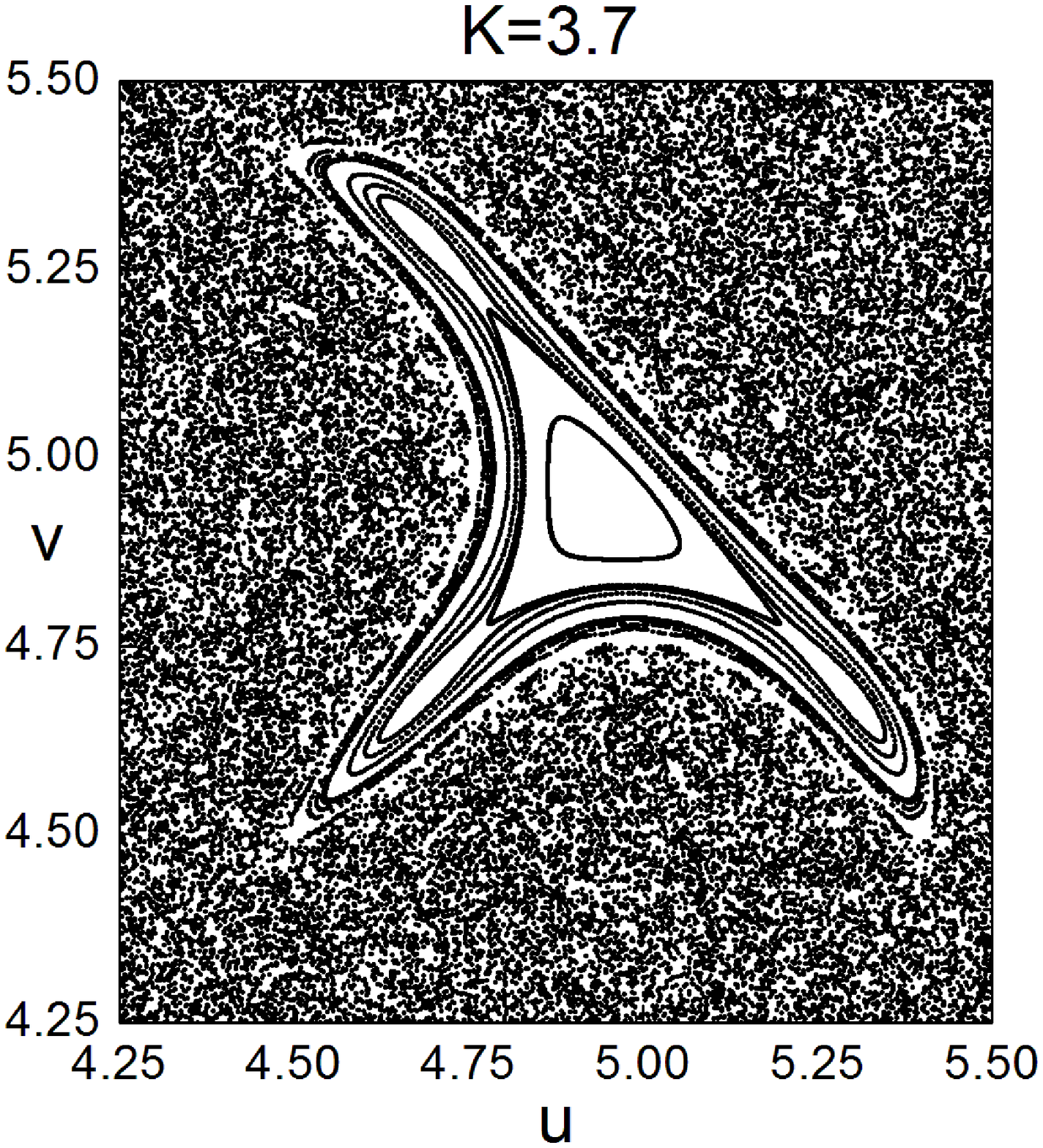}
\hspace{-0.4cm}
\includegraphics[height=5cm,angle=0,clip=]{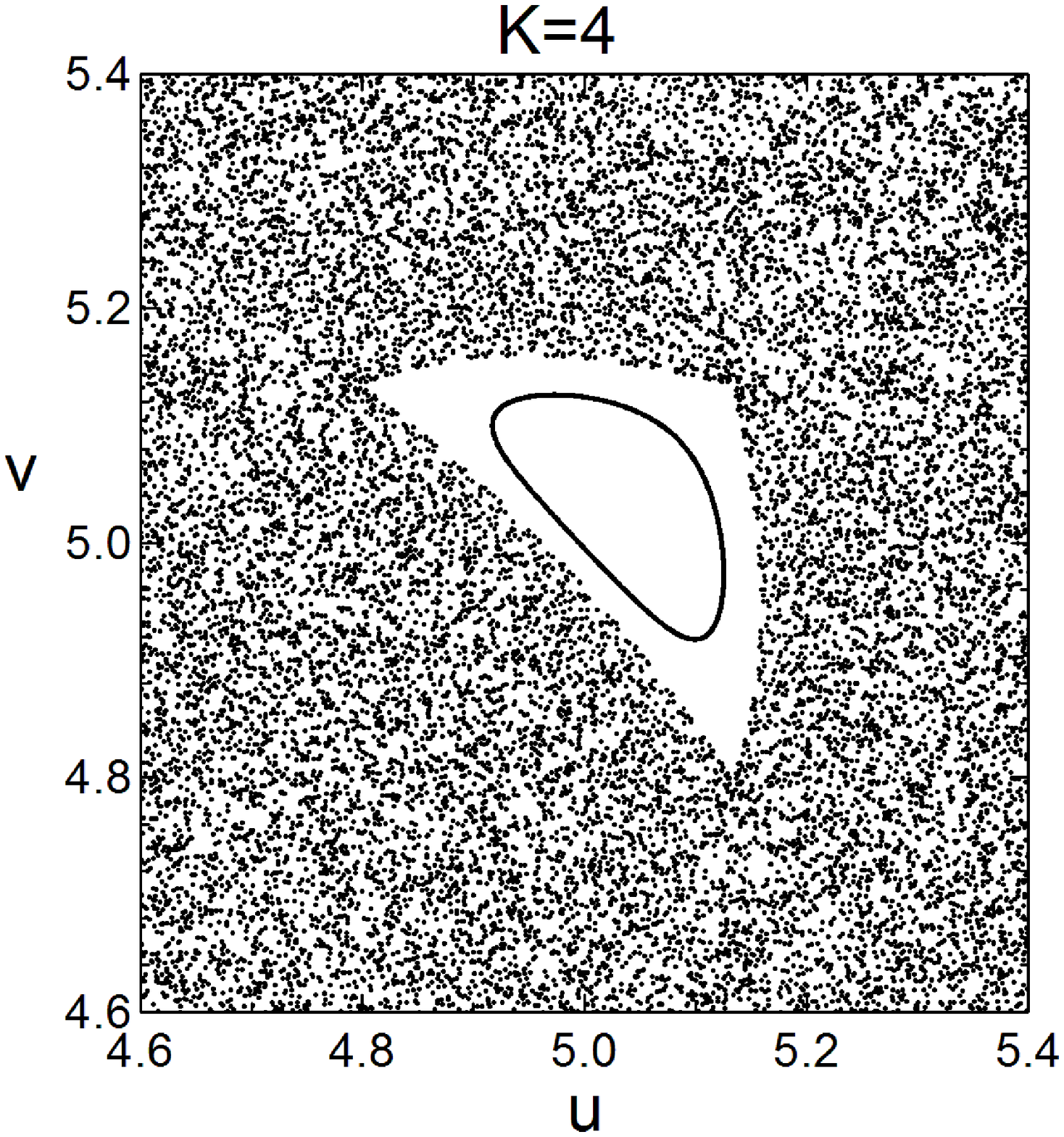}\\
\hspace{-0.3cm}
\includegraphics[height=4.5cm,angle=0,clip=]{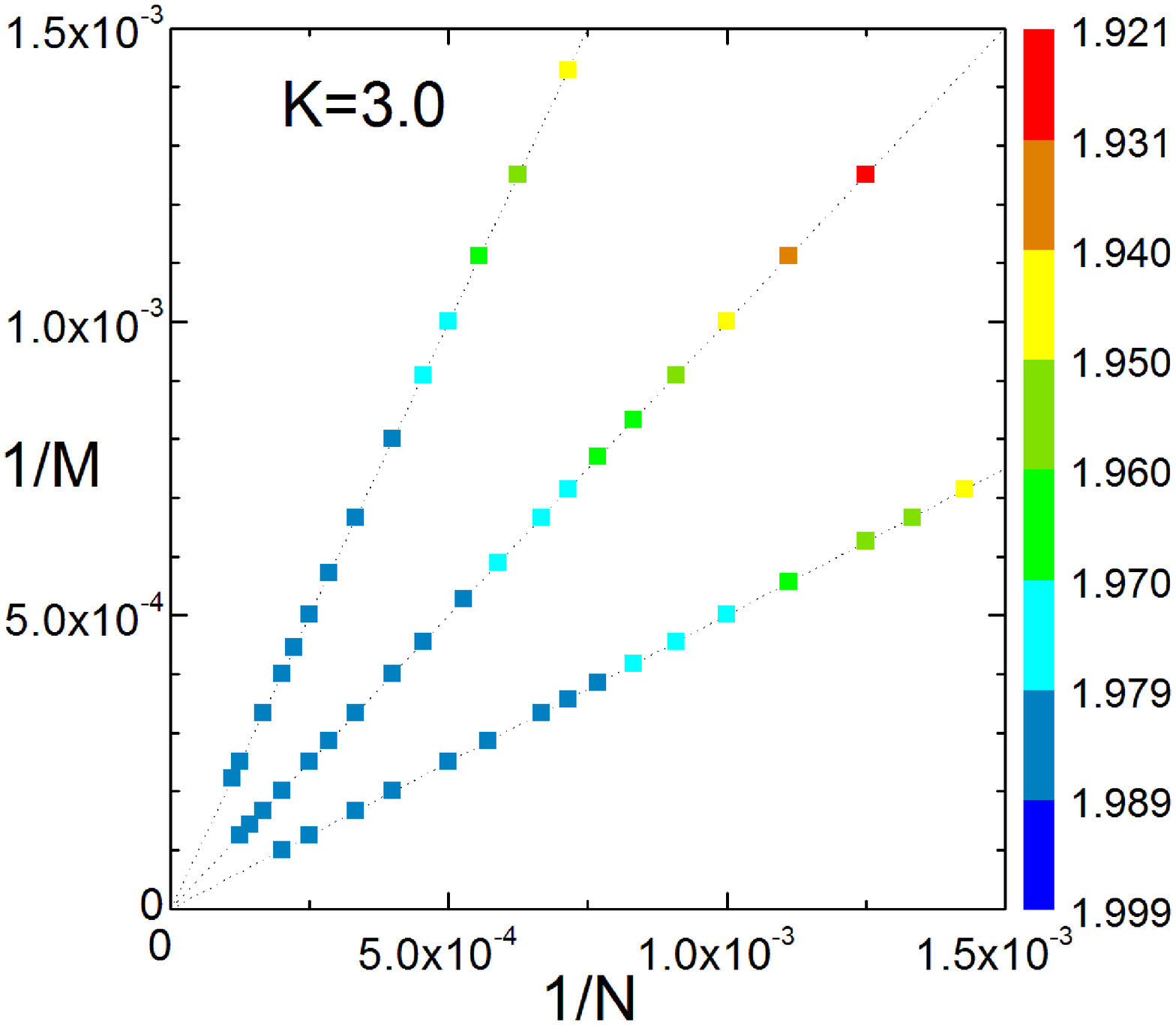}
\hspace{-1cm}
\includegraphics[height=4.5cm,angle=0,clip=]{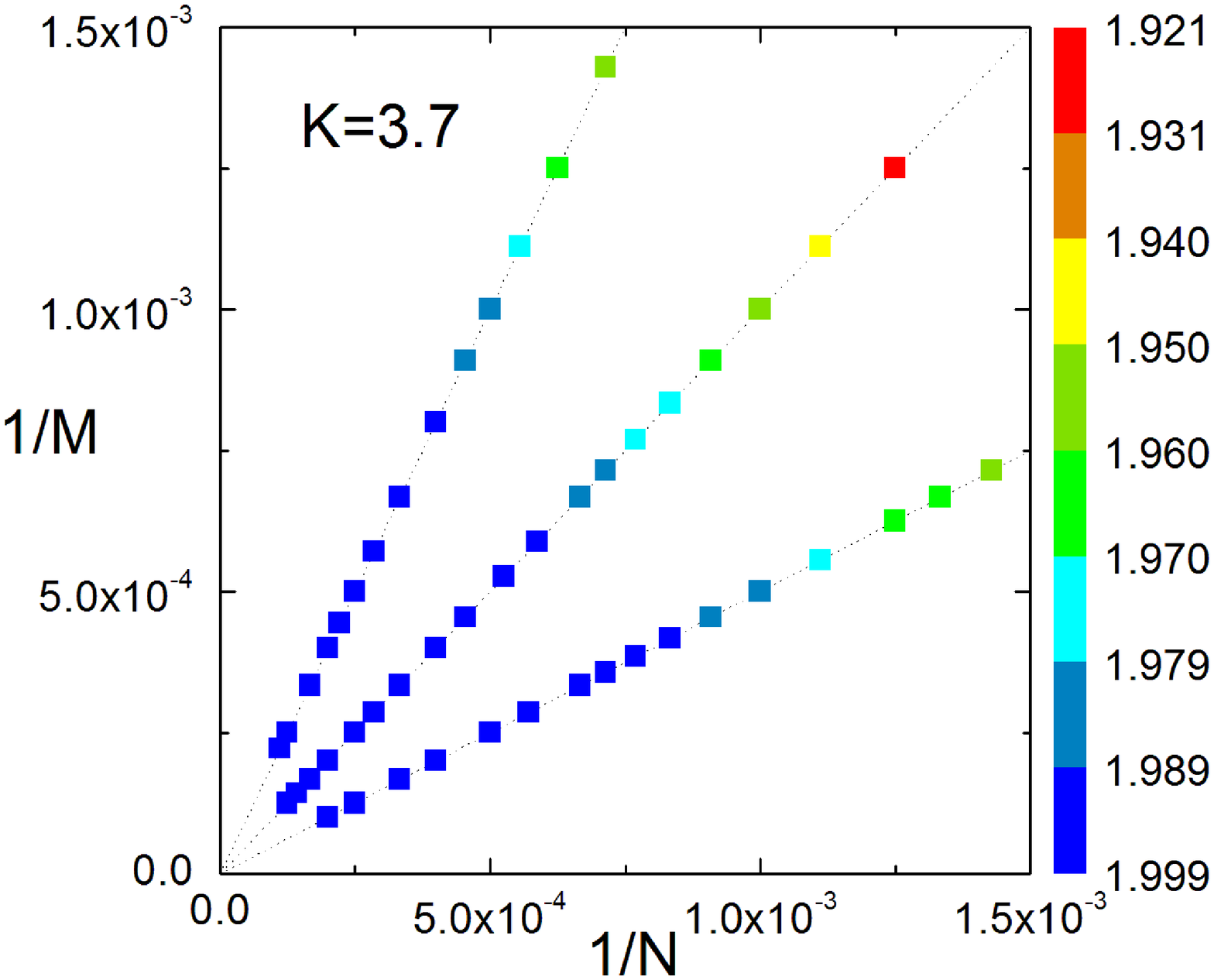}
\includegraphics[height=4.5cm,angle=0,clip=]{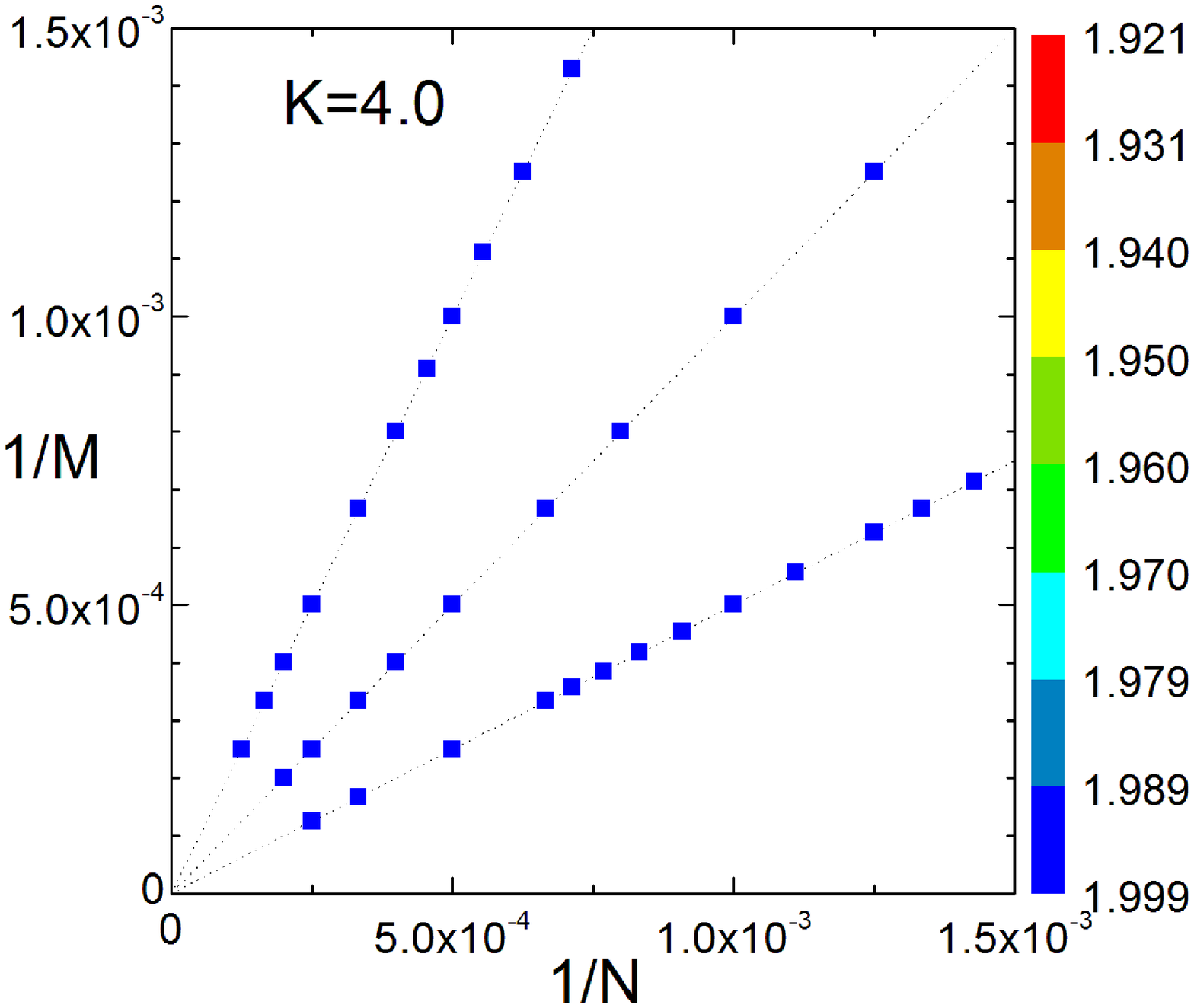}
\caption{\label{fig:fractalcolor}
{\small
(Color online)
Above: Details of the  phase portraits of the web map, for three representative values of the map parameter, $K$.
Below: Color maps of the box counting dimension of $M$ finite time  trajectories embedded inside their respective chaotic sea. $N$ is the number
of iterations; the  $M$ initial conditions have been randomly chosen all over the band $[2.7,3.8]\times[0,2 \pi]$.  }}
\end{figure}

We have observed that the values of $K$ where we detect such a  behavior appear to be related
to the existence in the phase space of a hierarchical organization of island chains of stability that
make the surrounding chaotic orbit to have structure at all scales, as we previously pointed out.
Consequently, the chaotic  trajectories can be characterized as (fat) fractals, and their fractal dimension appears to be $d_f<2$.
Fig.~\ref{fig:fractalcolor} shows, for some representative map parameter values, the color
maps of the box-counting  dimension   of  finite sets (i.e. $M<\infty$) and finite-time trajectories
(i.e. $N<\infty$) embedded inside the chaotic sea, together with a detail of their respective phase
portraits.
The box-counting fractal dimension has been calculated through a Matlab  implementation of the
Hou algorithm \cite{Souza2011, Hou76} that  optimizes memory storage and  significantly reduces
time requirements with respect to other box-counting calculation procedures.

\begin{figure}[h!]
\centering
\hspace{-0.5cm}
\includegraphics[height=6.5cm,angle=0,clip=]{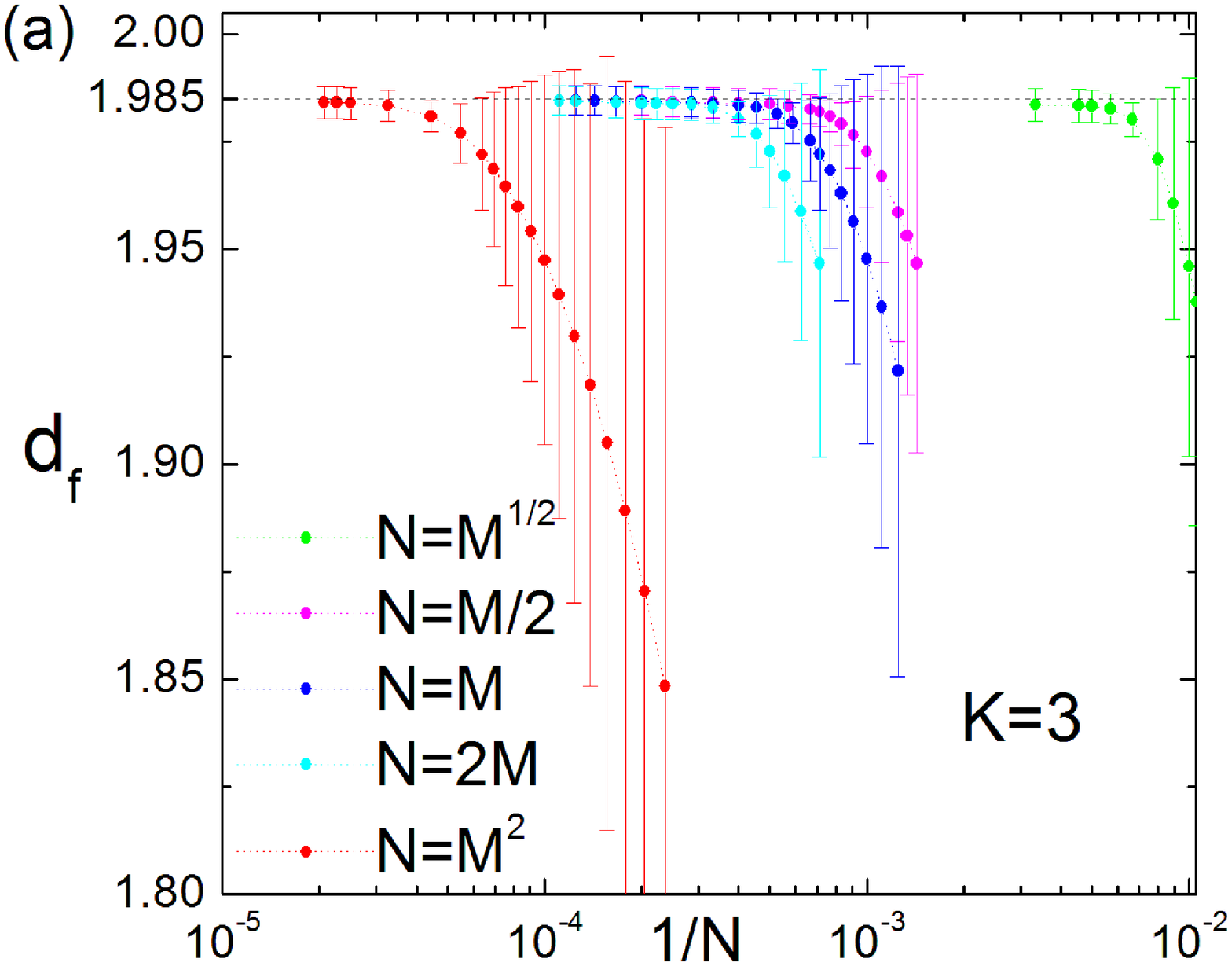}
\hspace{-0.5cm}
\includegraphics[height=6.5cm,angle=0,clip=]{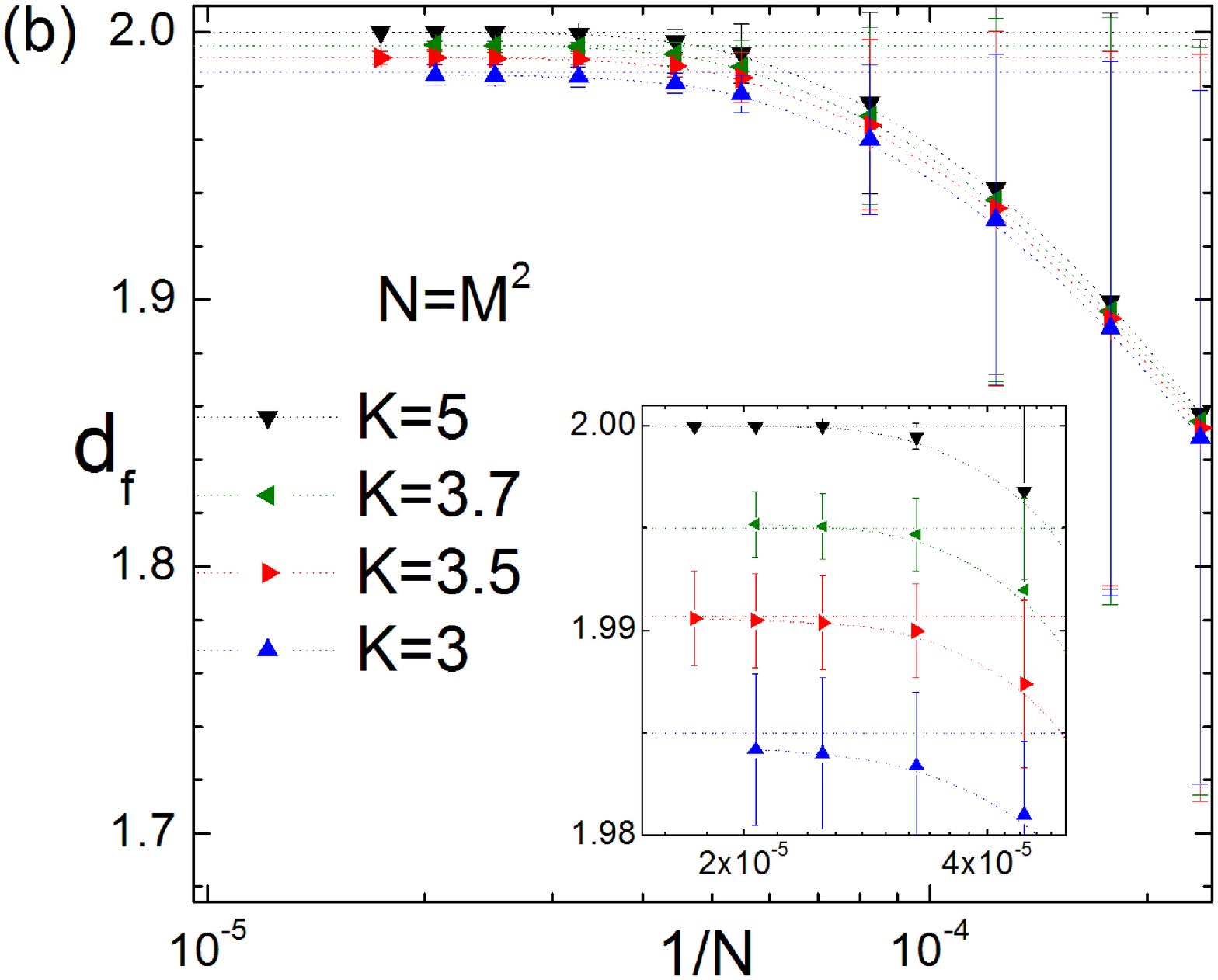}
\caption{\label{fig:fractaldimension}
{\small
(Color online) (a) Box counting dimension of the phase portrait  calculated through the first $N$ iterates of
$M$ trajectories in the $K=3$ web map. The initial conditions have been randomly chosen inside
the band $[2.7,3.8]\times[0,2 \pi]$ and, consequently, the  trajectories are embedded in the chaotic sea. In all cases, the $N\to \infty$ limit
appears to be $d_f = 1.985\pm 0.004<2$. (b) The same for some representative values of $K$, and $N=M^2$. }}
\end{figure}

The $N\to \infty$ limit probability distributions were expected to be Gaussians. With the aim of
characterizing the convergence, we have analyzed the $N\to \infty$  limit of the fractal dimension,
as $N\to \infty$ and $M\to \infty$, through different paths over the $(N,M)$ space.
Fig.~\ref{fig:fractaldimension}~(a) demonstrates, for $K=3$ that, no matter the path to be chosen, the
limit box-counting dimension is $d_f=1.985\pm 0.004<2$. Fig.~\ref{fig:fractaldimension}~(b) shows
that an analogous behavior has been found for other values of $K$
($K=3.5$, $K=3.7$,  \dots) that are associated to slow convergence behavior of the kind shown
in Fig.~\ref{fig:HistgK35}~(a). On the contrary, in the case of strongly chaotic phase space ($K=5$)
or fast convergence to Gaussian distribution ($K=3.8$ and $K=4$), the $N\to \infty$  box-counting
dimension is  $d_f = 2$.

\begin{figure}[h!]
\centering
\includegraphics[height=8cm,angle=0,clip=]{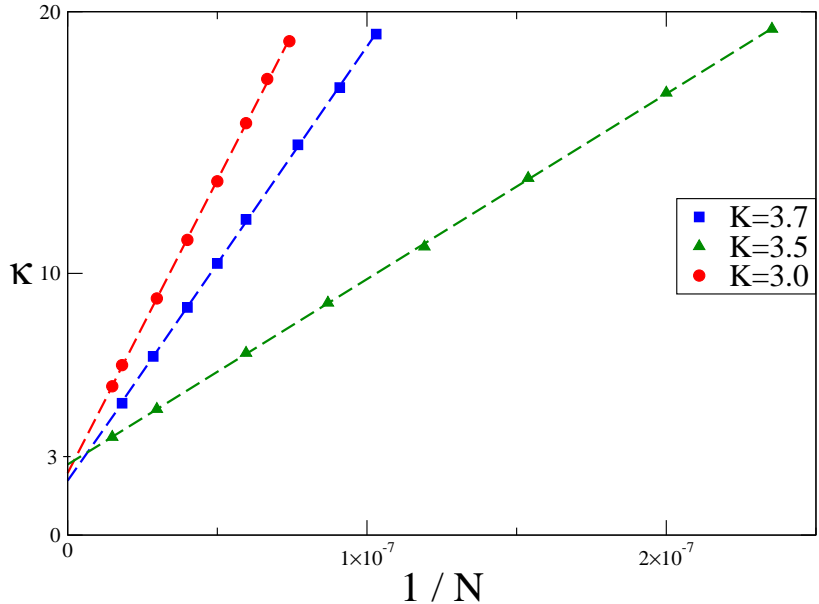}
\caption{\label{fig:kurtosisKs}
{\small
(Color online) Kurtosis of the probability distribution of the sums of iterates of finite-time
trajectories that are embedded inside the chaotic sea, for some representative values of the
parameter of the map that exhibit a fractal dimension of the chaotic sea $d_f<2$.
The asymptotic values of the kurtosis are calculated as $\kappa=2.37\pm0.06$ for
$K=3.$, $\kappa=2.70\pm0.04$ for $K=3.5$ and $\kappa=2.08\pm0.05$ for $K=3.7$ from
the figure. Regression coefficients are about $0.9999$.
Observe that the kurtosis does not appear to converge to the characteristic value of a Gaussian,
namely $\kappa_G=3$. Consequently, we infer that the $N\to \infty$ limit pdf is not a Gaussian.}}
\end{figure}

Let us now estimate the  kurtosis of the $N\to \infty$ limit pdfs, by calculating the kurtosis of the
finite $N$ pdfs, $\kappa$. We will make use of the expression:
\begin{equation}
\label{eq:kappa}
\kappa=\frac{1}{\sigma^4}
\left[\frac{1}{M}\sum_{i=1}^M(y_i-\mu)^4\right]
\end{equation}
where $y_i$ ($i=1, \dots, M$) are the sums of the $N$ iterates in Eq.~(\ref{eq:variable}), $\mu$ is the
arithmetical average of the  sums, and $\sigma$ is the variance.
Fig.~\ref{fig:kurtosisKs} shows, for some representative values of $K$ that drive to a fractal dimension
of the chaotic sea ($d_f<2$), that the $N\to \infty  $ limit distribution does not in fact appear to
converge to a Gaussian, i.e., $\lim_{N\to \infty}\kappa\ne 3$. On the contrary, the values of $K$
that drive to a chaotic sea with $d_f=2$, verify $\lim_{N \to \infty}\kappa= 3$.
Table~\ref{table1} shows, for typical values of $K$, the estimated  ($N\to \infty $) limit value of the
kurtosis of the limit pdf,  and the estimated values of ($N\to \infty $) limit of  fractal dimension in a
set of ($M\to \infty$) trajectories. We conclude that a non trivial dependence between $\kappa$
and $d_f$ exists. Indeed the  fat fractal  dimension embedded inside the chaotic sea converges slowly
 towards some generalized distributions  whose  long tails preclude  a Gaussian
characterization.

\begin{table}
\centering
\begin{tabular}{|c |c|c| }
  \hline
  $K$&$\kappa$ & $d_f$ \\
  \hline
  \hline
$ 3$ & $2.37\pm 0.06$ & $1.9846\pm0.0034$ \\
$3.5$ &$2.70\pm 0.04$ &$1.9909\pm 0.0020$\\
$3.7$ & $2.08\pm 0.05$&$1.9963\pm 0.0014$ \\
$3.8$ &$3$ & $2$\\
$4$ &  $3$ & $1.9989\pm0.0004$\\
$5$    & $3$ &$2$\\
  \hline
\end{tabular}
\caption{\label{table1}
{\small
Estimated  ($N\to \infty $) limit value of the kurtosis of the limit pdf,  and the estimated values of
($N\to \infty $) limit of  fractal dimension in a set of ($M\to \infty$) trajectories of typical values of
the web map parameter.}}
\end{table}

\section{$q$-statistical indices}
\label{sec:4}

The previous section deals with the  stationary and quasi stationary characterization of the web map. As we pointed out, in some cases, the finite $N$ probability distribution functions can be properly fitted with  a $q$-Gaussian that is characterized by a $q_{stat}$ index. Let us now make a review of other new  results related to other $q$-statistical indices, that have been  obtained for the web map, and are consistent with the results that were obtained in the standard map  \cite{Ruiz}.

For many complex systems, the sensitivity to initial conditions, $\xi(t)$ is described
by a generalized function, $e_q(x)=[1+(1-q)x]^{\frac{1}{1-q}}$ (with $e_1(x)=\exp(x)$), referred to as
$q$--exponential  \cite{Plastino97}. More precisely,

\begin{equation}
\small{ \xi(t)\equiv\lim_{\|{\bf \Delta x}(0)\|  \to {\bf 0}}\frac{\|{\bf \Delta x}(t)\|}{\|{\bf \Delta x}(0)\|}=
\left[1+(1-q_{sen})\lambda_{q_{sen}}t\right]^{\frac{1}{1-q_{sen}}}\equiv e_{q_{sen}}^{\lambda_{q_{sen}} t}},
\label{eq:qsensitiv}
\end{equation}
${\bf\Delta x}(t)$ being the temporal dependence of the discrepancy between two very close initial conditions at
time $t$, where $q_{sen}$ and $\lambda_{q_{sen}}$ ({\it generalized Lyapunov coefficient}) are parameters.
When ergodic behavior dominates over the whole phase space,
Eq.~(\ref{eq:qsensitiv})
 recovers the standard
exponential dependence $\xi(t)=e^{\lambda t}$ ($q_{sen}=1$ implies
$\lambda_{q_{sen}}\to \lambda\equiv \lambda_1$, where $\lambda$ denotes the standard Lyapunov exponent).

\begin{figure}[h!]
\centering
\hspace{-0.5cm}
\includegraphics[height=5cm,angle=0]{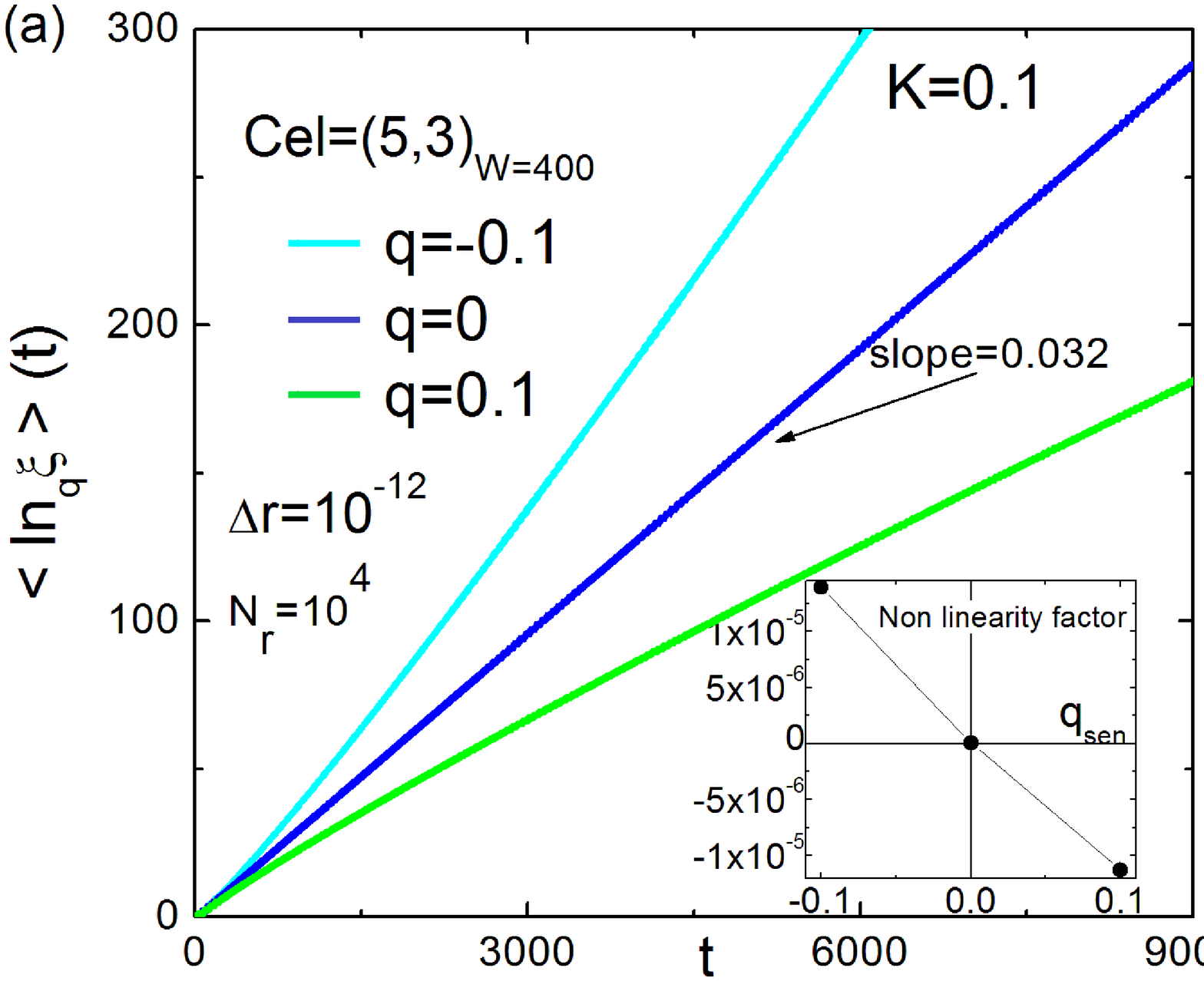}\hspace{0.3cm}
\includegraphics[height=5cm,angle=0]{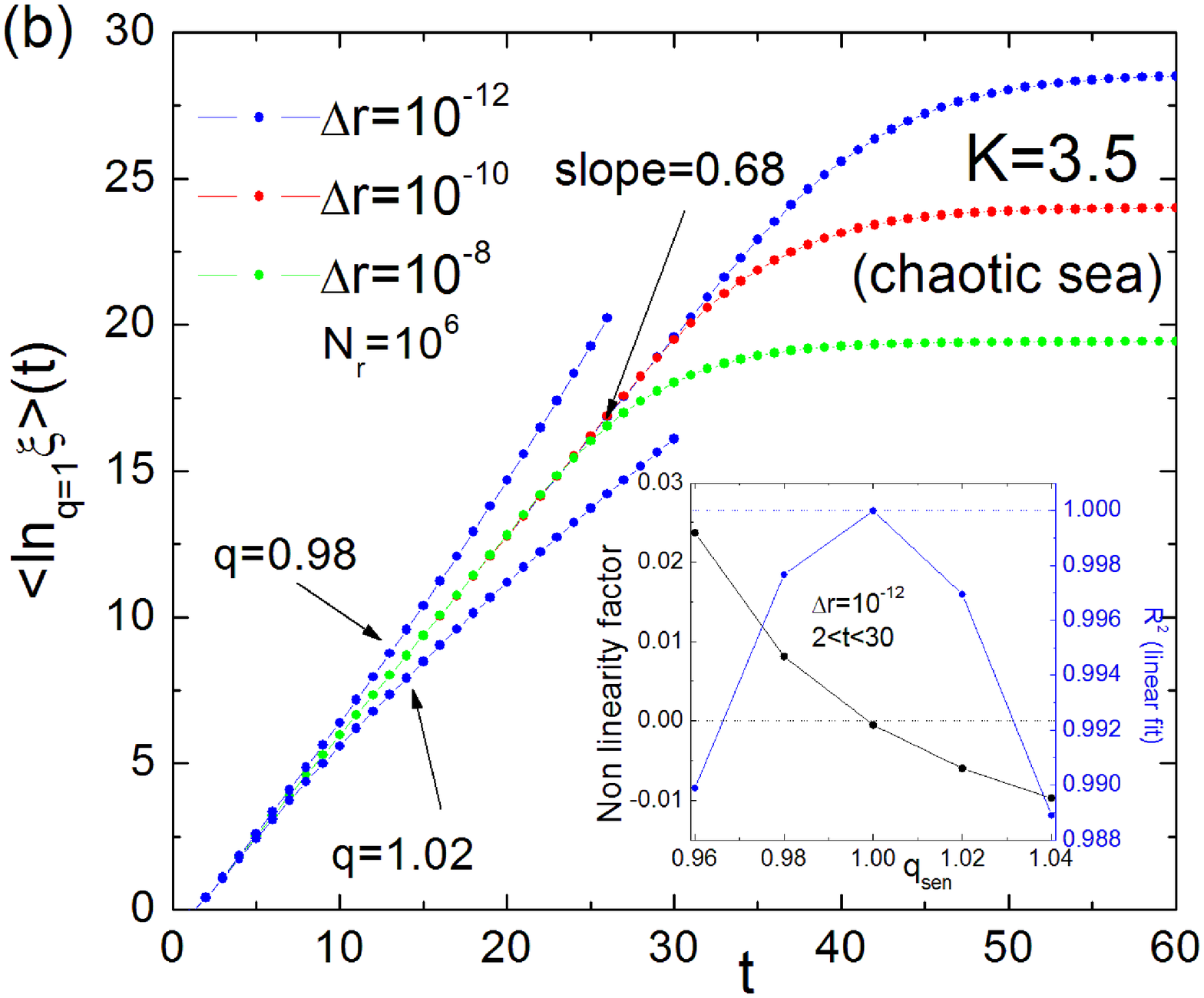}\hspace{-0.5cm}
\includegraphics[height=5cm,angle=0]{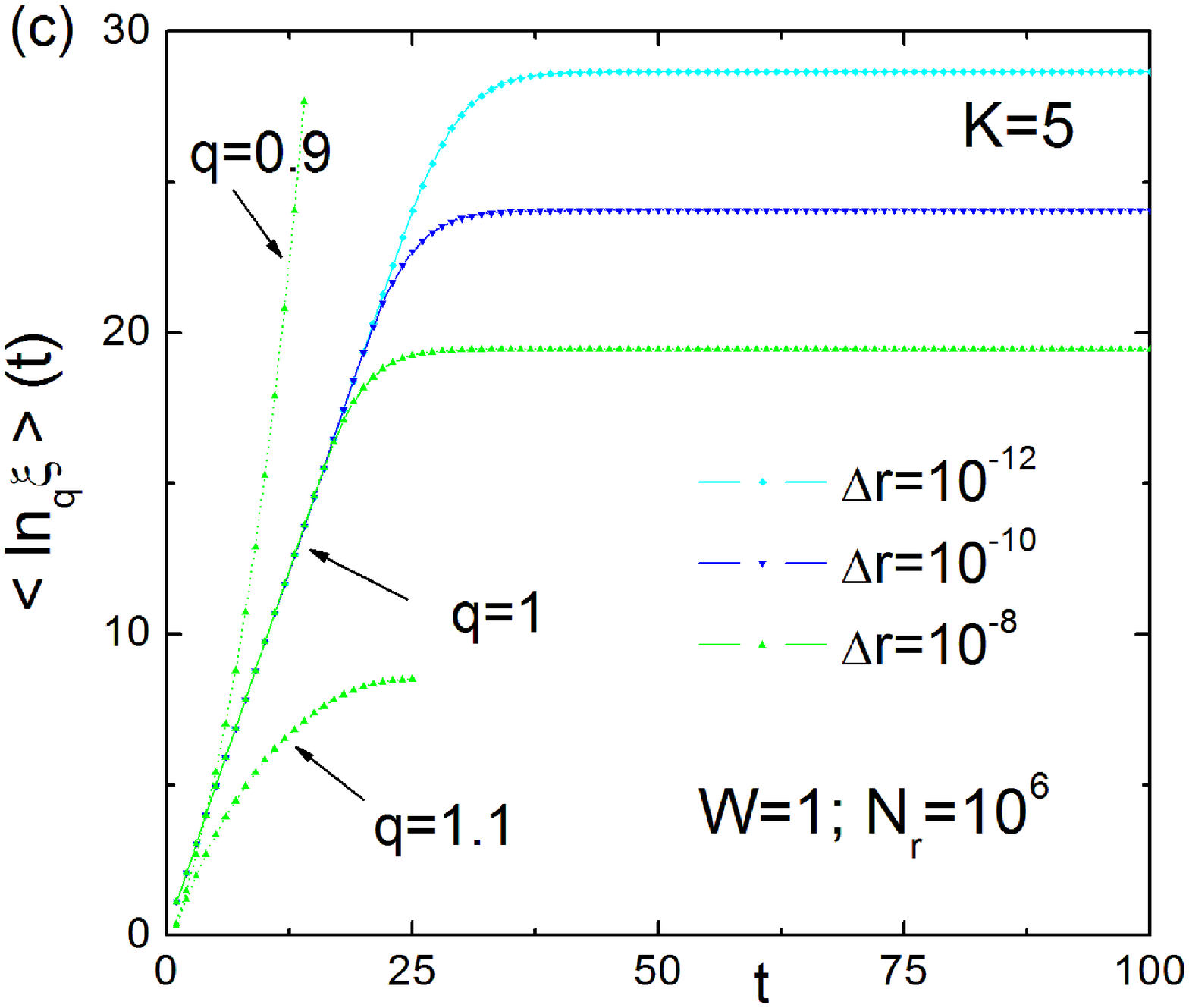}\hspace{-0.5cm}
\vspace{-0.4cm}
\caption{\label{fig:xiks}{\small
(Color online) Average of the $q$-logarithm of the sensitivity to initial conditions over $N_r$ realizations
for (a)~the $K=0.1$ web map dominated by the regular islands, (b)~the chaotic sea of the $K=3.5$ web map,
and (c)~the $K=5$ web map dominated by the chaotic sea. $\Delta r\equiv {\bf \Delta x}(0)$ is the distance between two initial conditions
that have been randomly chosen within a particular cell of the $W$-partitioned phase portrait (see text).
Insets  demonstrate that (a)~$q_{sen}=0$ and $\lambda_0=0.032$  inside the
$(5^{th},3^{th})$ cell of $W=20\times 20$ equally partitioned $K=0.1$ phase space,
and (b)~$q_{sen}=1$ and $\lambda_1=0.68$ inside the central cell of the $W=3\times 3$
equally partitioned phase space. (c) The  $K=5$ web map satisfies
$q_{sen}=1$ and $\lambda_1=0.96$.  }}
\end{figure}

This description is verified for both strongly and weakly chaotic regimes.
Fig.~\ref{fig:xiks}  shows, for different values of $K$ parameter and different values of $q$, the average
of $\ln_q\xi (t)$ (where $\ln_qx\equiv (x^{1-q}-1)/(1-q)$ is the inverse function of the $q$-exponential, and
$\ln_1 x=\ln x$) over $N_r$ realizations. Each realization starts with a randomly chosen pair of very close
initial conditions, localized inside a particular cell of the $W$ equally partitioned phase space. We considered
decreasing initial discrepancies in Eq.~(\ref{eq:qsensitiv})
 (see Fig.~\ref{fig:xiks}~(b-c)), so as to obtain a well
defined behavior for increasingly long times, and verify a nontrivial property \cite{Ananos}, namely that a
special value of $q$ exists, noted $q^{av}_{sen}$ (where  $av$ stands  for  {\it average}), which yields a
linear dependence of  $\langle\ln_q\xi \rangle$ with time. In other words, we verify
$\langle \ln_q \xi(t)\rangle\approx \lambda_{q_{sen}^{av}} t$, where the linear coefficient
$\lambda_{q_{sen}^{av}}$ is a $q$-generalized Lyapunov coefficient. When stability islands dominate the
phase space ($K=0.1$), Fig.~\ref{fig:xiks}~(a) exhibits that $q_{sen}^{av}=0$, and  the generalized Lyapunov exponent
$\lambda_{q_{sen}^{av}}=0.032\pm 0.001$ (time steps)$^{-1}$ characterizes the local sensitivity to initial conditions.
In the  chaotic sea, e.g., Fig.~\ref{fig:xiks}~(b) for the trajectories embedded inside the chaotic sea of the $K=3.5$ map, and
Fig.~\ref{fig:xiks}~(c) for the strongly chaotic $K=5$ map, where an exponential sensitivity to initial conditions is verified
as $q_{sen}^{av}=1$, and the slope of intermediate regimes demonstrate to be
$\lambda_{q_{sen}^{av}=1}=0.68\pm 0.01 $ (time steps)$^{-1}$ and $\lambda_{q_{sen}^{av}=1}=0.96\pm 0.01$
(time steps)$^{-1}$, respectively. The proper $q$ indices are obtained by fitting the curves with the
polynomial $A+Bt+Ct^2$ over the intermediate regime (before saturation), and comparing their nonlinearity
measure $R\equiv C/B$. The optimum value of the entropic index corresponds to $R=0$ (a straight line).
The intermediate regime that we consider is such that the linear regression coefficient typically is $0.9999$.

All these results are consistent with the  sensitivity to initial conditions behavior of the standard map \cite{Ruiz}. However, it must be
said that the sudden jump we found from $q_{sen}=0$ to $q_{sen}=1$ in the frontier from regular to strongly chaotic
regions is not the common rule. In fact, other dynamical systems exist that present, at the frontier
between the chaotic and regular regions, a $q$-exponential sensitivity to initial conditions behavior, where
$q$ monotonically increases from zero to unity when the nonlinear parameter  increases from zero to a
critical value for which the phase space is fully chaotic. Such is the case, for example, of the quantum kicked
top and the classical kicked top map  \cite{Weinstein,Queiros}.

With respect to the entropy production per unit time, we may conveniently use the $q$-generalized entropy
($k=1$, henceforth) \cite{Tsallis, Tsallis2010}

\begin{equation}
{\small S_q(t)=\frac{1-\sum_{i=1}^W p_i^q}{q-1} \qquad
\left(S_1=S_{BG}\equiv - \sum_{i=1}^W p_i \ln p_i\right)} \,,
\label{eq:qentropy}
\end{equation}
and one special value of the entropic index $q$, noted $q_{ent}$ ($ent$ stands for {\it entropy}), makes the
entropy production to be finite. The Boltzmann-Gibbs entropy, $S_{BG}$, is expected to be the appropriate one
when the LLE is definitively positive and the phase space is dominated by a strongly chaotic sea, i.e.,
$q_{ent}=1$ (hence $S_{q_{ent}}\to S_{BG}\equiv S_1$). In other cases $q_{ent}\ne 1$. To be more precise, the $q$-entropy
production is estimated by dividing the phase space in $W$ equal cells and randomly choosing $N_{ic} \gg W$
initial conditions inside one of the $W$ cells (typically $N_{ic}=10 W$). We follow the spread of
points within the phase space, and calculate
Eq.~(\ref{eq:qentropy})  from the set of occupancy probabilities
$\{p_i(t)\}$ ($i=1,2,\dots,W$). We repeat the operation $N_{c}$ times, choosing different initial cells within
which the $N_{ic}$ initial conditions are chosen, and we finally average $S_q(t)$ over the $N_{c}$ realizations,
so as to reduce fluctuations. The proper value of the entropic parameter $q_{ent}^{av}$ is the special value
of $q$ which makes the averaged $q$-entropy production per unit time to be finite. The $q$-entropy production per
unit time

\begin{equation}
s_{q_{ent}^{av}}(t)=\lim_{t \to \infty }\lim_{W \to \infty }\lim_{N_{ic} \to \infty }\frac{\langle
S_{q_{ent}^{av}}\rangle_{N_{c}}}{t}\label{eq:qKolmogorov}
\end{equation}
is calculated taking into account that the partitions of phase space must be such as to obtain
robust results.

\begin{figure}[h]
\centering
\includegraphics[height=6.5cm,angle=0,clip=]{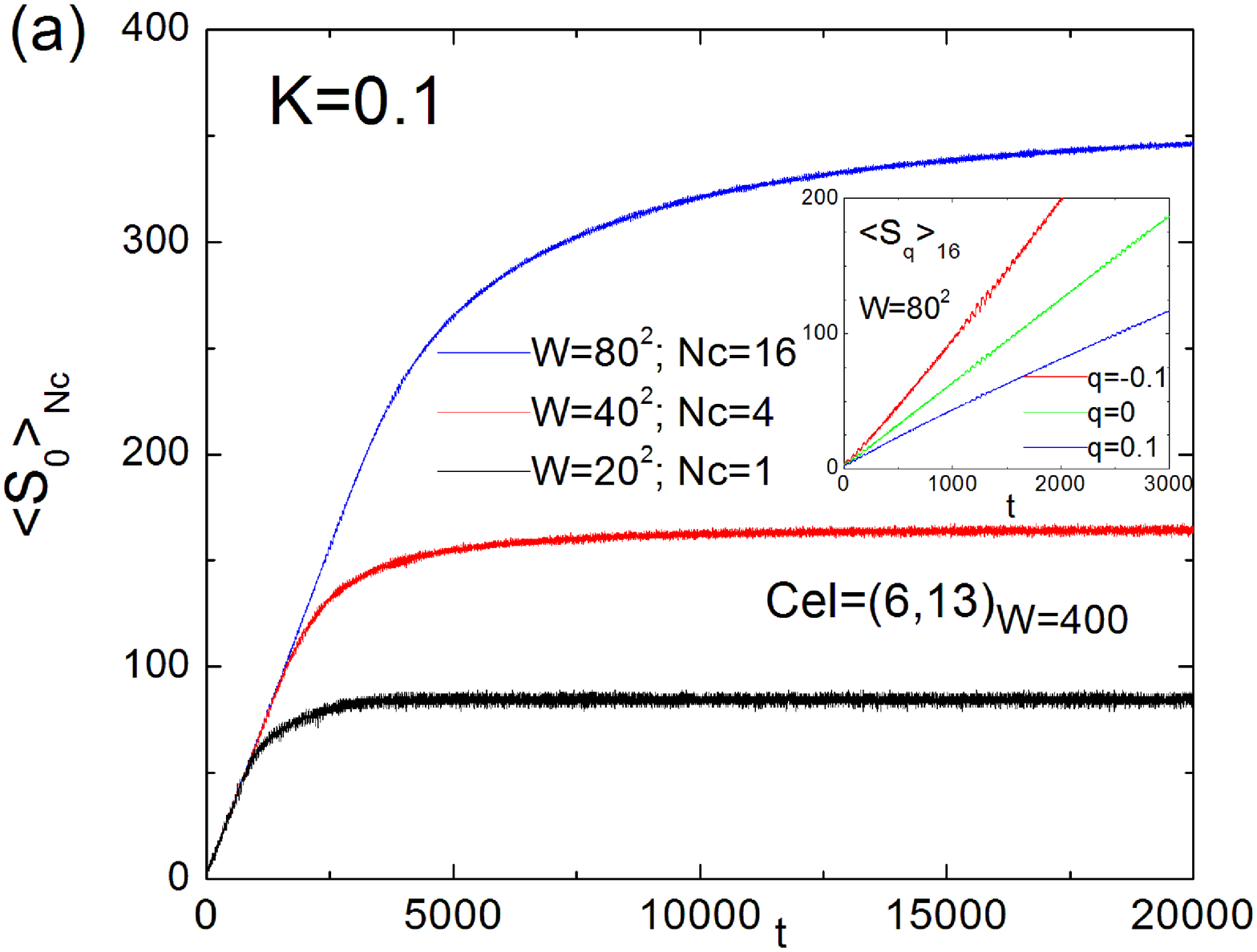}
\includegraphics[height=6.5cm,angle=0,clip=]{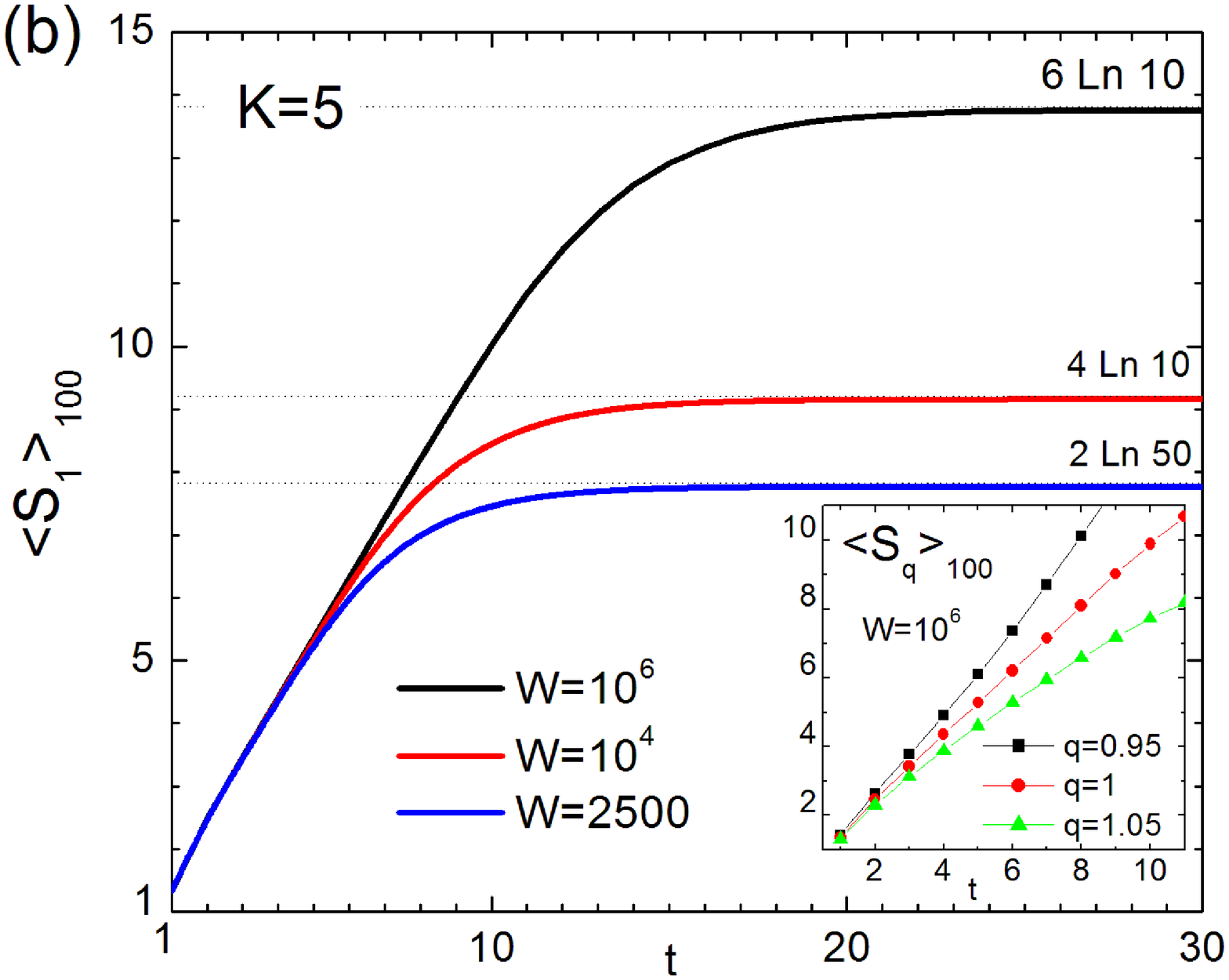}\vspace{-0.4cm}
\caption{\label{fig:Sq}
{\small (Color online) $(q=0)$-Entropy production as a function of time, for (a)~a particular cell in
the partitioned phase space of $K=0.1$ web map, and (b)~the whole phase space of $K=5$ web
map, where the slope in the intermediate temporal regime is $0.96$. The insets demonstrate that
the proper values of $q$ are, respectively, $q=0$ (for $K=0.1$) and $q=1$ (for $K=5$). $N_c$ represents
the number of averaging cells.}}
\end{figure}

When stability islands dominate the phase space (e.g., $K=0.1$), $q_{ent}^{av}=0$, as shown in
Fig.~\ref{fig:Sq}~(a). Fluctuations of the entropy production for a particular cell of a given coarse
graining $W$ (e.g.,  $(6^{th},13^{th})$ cell of a $W=20\times 20$ equipartitioned phase space)
can be reduced by taking a thinner coarse graining (i.e., by increasing $W$) and averaging over
the cells of this new partition that exactly fill the original one (in our example, averaging over the
new cells which the $(6^{th},13^{th})$ cell was divided into). The BG entropy is, as expected, the
appropriate one for the $K=5$ strongly chaotic case ($q_{ent}^{av}=1$). In fact, for the strongly chaotic regime,  $q$-entropy production per unit time must satisfy
$s_{q_{ent}^{av}=1}=\lambda_{q_{sen}^{av}=1}$. In particular, for the  $K=5$ web map, $s_{q_{ent}^{av}=1}=\lambda_{q_{sen}^{av}=1}=0.96\pm 0.01$
(time steps)$^{-1}$ as shown in Fig.~\ref{fig:Sq}~(b). These numerical results are analogous to
those obtained for the standard map in \cite{Ruiz}. However, in the case of intermediate values of
 $K$  where the chaotic sea and the stability islands coexist, the intermediate regime of $q$-entropy production that
satisfies $s_{q_{ent}^{av}=1}=\lambda_{q_{sen}^{av}=1}$ is attained for much longer times and
much finer partitions. A high computational effort is required to numerically obtain, for $K=3.5$
web map, that $s_{q_{ent}^{av}=1}=\lambda_{q_{sen}^{av}=1}=0.68$. The need of a finer
partitioned phase space is in fact the first feature  that we have found that distinguishes the coarse
graining behavior of the  web map with respect to the standard map.

\begin{figure}[h!]
\centering
\includegraphics[height=8cm,angle=0,clip=]{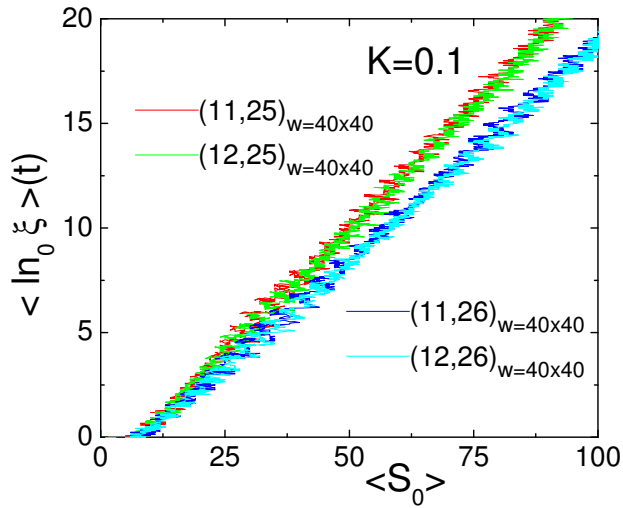}\vspace{-0.4cm}
\caption{\label{fig:GSqPhi}
{\small (Color online) $(q=0)$-logarithm of sensitivity to initial conditions, as a function of the
$(q=0)$-entropy evolution of the $(6^{th},13^{th})$ cell in the $W=20\times20$ partitioned
space of the $K=0.1$ web map. The $W=2\times 2$ partition inside that
$(6^{th},13^{th})_{20\times 20}$ cell, splits the linear behavior, as their respective slopes
characterize the local sensitivity to initial conditions.
These numerical results have been obtained for ${\bf \Delta x}(0)=10^{12}$ and $N_r=10^5$.
}}
\end{figure}

The strong fluctuations that characterize weak chaos entropy production can also be overcome
analyzing $S_q(t)$ as a function of $\langle \xi(t) \rangle $ in a particular cell. Observe that
Fig.~\ref{fig:GSqPhi} shows that the bound values of $(q=0)$--entropy are related to
$(q=0)$--sensitivity as $\langle S_{q_{ent}^{av}=0}\rangle\propto  \ln_0\langle \xi(t) \rangle$ but
the proportionality factor is not equal one, i.e., $s_{q_{ent}^{av}=0}\ne \lambda_{q_{sen}^{av}=0}$.
The cells of a partitioned cell exhibit different slopes, that  characterize the local sensitivity to initial
conditions. This is the same behavior that has been  found in the weak chaos regime of the
standard map \cite{Ruiz}.

\section*{Conclusions}
\label{sec:5}

Some previous results found on the standard map \cite{Ugur,Ruiz}
are also displayed by the web map. For instance,
when the phase space is  filled by a chaotic sea,
i.e., LLE $\gg 0$, Gaussian distributions emerge quickly.
When the phase space is dominated by stability islands
(LLE $\sim 0$),
the distributions converge to a $q$-Gaussian. The persistence of the value $q_{stat}=1.935$ for both standard and web systems
constitutes a remarkable result, pointing towards universality. With respect to the analysis of other indices, namely $q_{sen}$ and $q_{ent}$
indicate two distinct cases,
$q_{sen} = q_{ent} = 1$ for the strongly chaotic regions,
and
$q_{sen} = q_{ent} = 0$ for the regions with regular islands.

The behavior of these  two maps departs from each other
when the phase space displays islands embedded within a chaotic sea.
In fact, for the standard map, the coexistence of these two regimes leads to
a simple superposition of the probability distributions for each case.
 On the contrary, in the web map, the coexistence of both regimes induces, in case of a significant
 sticky behavior, a statistical behavior  which differs from the one that was detected for the standard map. Indeed,
the central part of the distributions in the chaotic sea slowly evolves towards a Gaussian,
but neat non-Gaussian tails persist. This fact is related to the  fractal dimension of finite sets and
finite-time trajectories embedded inside the chaotic sea, $d_f < 2$, which is consistent
with \cite{Benettin86}. But even in the limit of infinite-time trajectories, kurtosis $\kappa$ of these
distributions yields $\kappa < 3$  in the case of significant sticky behavior  --- which suggests that the
distributions will not converge to Gaussians --- and $d_f < 2$. The existence of non Gaussian limit
distributions in a fractal support appears to be an interesting finding that can be related with the analytical
results of Carati \cite{Carati}, who characterized the fractal dimension of the orbits compatible with
temporal averages of dynamical variables in a $q$-statistical scenario.
%

\section*{Acknowledgments}
 One of us (G. R.)  thanks L. J. L. Cirto for fruitful discussions about various computational
 problems.
This work has been partially supported by CNPq and Faperj (Brazilian Agencies), and by
TUBITAK (Turkish Agency) under the Research Project number 115F492.
GR, UT and CT also acknowledge partial financial support by the John Templeton
Foundation (USA).


\end{document}